%% file: SUN_Cooling.tex
\documentclass[aps,twocolumn,showpacs,pra,superscriptaddress,floatfix,10pt]{revtex4-1}

\usepackage{graphicx,amsmath, amssymb, amsthm, amsfonts, mathtools}
\usepackage[usenames,dvipsnames]{xcolor}
\usepackage{hyperref}			   
\usepackage{braket}
\usepackage{xcolor} 

\def\beq{\begin{equation}}
\def\eeq{\end{equation}}
\def\beqa{\begin{eqnarray}}
\def\eeqa{\end{eqnarray}}

\begin{document}

\title{State selective cooling of $\mathrm{SU}(N)$ Fermi-gases}

\author{Aaron Merlin M\"{u}ller} 
\affiliation{Institute for Theoretical Physics, ETH Z\"{u}rich, 8093 Z\"{u}rich, Switzerland}
\affiliation{Institute of Physics, \'{E}cole Polytechnique F\'{e}d\'{e}rale de Lausanne (EPFL), CH-1015 Lausanne, Switzerland}

\author{Mikl\'{o}s Lajk\'{o}} 
\affiliation{Institute of Physics, \'{E}cole Polytechnique F\'{e}d\'{e}rale de Lausanne (EPFL), CH-1015 Lausanne, Switzerland}

\author{Florian Schreck} 
\affiliation{Van der Waals-Zeeman Institute, Institute of Physics, University of Amsterdam, Science Park 904, 1098 XH Amsterdam, the Netherlands}
\affiliation{QuSoft, Science Park 123, 1098 XG Amsterdam, the Netherlands}

\author{Fr\'{e}d\'{e}ric Mila} 
\affiliation{Institute of Physics, \'{E}cole Polytechnique F\'{e}d\'{e}rale de Lausanne (EPFL), CH-1015 Lausanne, Switzerland}

\author{Ji\v{r}\'{i} Min\'{a}\v{r}} 
\affiliation{Institute for Theoretical Physics, University of Amsterdam, Science Park 904, 1098 XH Amsterdam, the Netherlands}
\affiliation{QuSoft, Science Park 123, 1098 XG Amsterdam, the Netherlands}

\date{\today}


\begin{abstract}

We investigate a species selective cooling process of a trapped $\mathrm{SU}(N)$ Fermi gas using entropy redistribution during adiabatic loading of an optical lattice. Using high-temperature expansion of the Hubbard model, we show that when a subset $N_A < N$ of the single-atom levels experiences a stronger trapping potential in a certain region of space, the dimple, it leads to improvement in cooling as compared to a $\mathrm{SU}(N_A)$ Fermi gas only. We show that optimal performance is achieved when all atomic levels experience the same potential outside the dimple and we quantify the cooling for various $N_A$ by evaluating the dependence of the final entropy densities and temperatures as functions of the initial entropy. 
Furthermore, considering ${}^{87}{\rm Sr}$ and ${}^{173}{\rm Yb}$ for specificity, we provide a quantitative discussion of how the state selective trapping can be achieved with readily available experimental techniques.
\end{abstract}

\maketitle

\section{Introduction}

In recent years, there has been a considerable effort in experimental control of ultracold Fermi gases with the aim of realizing models of strongly interacting electrons, in particular the Hubbard model, upon loading the atoms into a deep optical lattice \cite{Esslinger_2010}. 
Of particular interest are ultracold quantum degenerate Fermi gases with nuclear spin $I$ that is decoupled from the electronic spin, such as ${}^{173}$Yb \cite{Fukuhara_2007_PRL, Taie_2010_PRL, sugawa2013ultracold} or ${}^{87}$Sr \cite{DeSalvo_2010_PRL, Stellmer_2013_PRA, stellmer2014degenerate}, which feature $N=2I+1$ hyperfine states in the ground state manifold.
The ${\rm SU}(N)$ Fermi gases have attracted considerable attention as they allow for $\mathrm{SU}(N)$ generalizations of the Hubbard model \cite{cazalilla2014ultracold} and can host a plethora of exotic phases including various spin orders and liquids \cite{wang2014competing,barbarino2015magnetic,chen2016synthetic,capponi2016phases,jen2018spin,Chung_2019_PRB}, Mott insulator-metal transitions and crossovers \cite{blumer2013mott,xu2018interaction}, valence bond solids and semimetals \cite{zhou2017finite,lang2013dimerized}, unconventional superconductors \cite{Wolf_2018_PRB} or collective motional modes \cite{choudhury2020collective}. Remarkably, some of these scenarios have been probed also experimentally for $N>2$ \cite{pagano2014one,PhysRevX.6.021030, Ozawa_2018_PRL, taie2020observation, Sonderhouse2020}.
The limit of large interaction gives rise to $\mathrm{SU}(N)$ magnetism \cite{Gorshkov_2010_NatPhys,Manmana_2011_PRA}, where the system can be effectively described in terms of a Heisenberg model. This stimulated theoretical investigations using representation theory \cite{nataf2014exact, nataf2016exact, kim2017linear, nataf2018density}, variational approaches \cite{dufour2015variational} or large scale simulations at finite temperature \cite{romen2020structure}. Furthermore, depending on $N$ and the lattice geometry, the Heisenberg Hamiltonians can be linked to Wess-Zumino-Witten models when at a critical point \cite{chen2015quantum,d2015renyi}, feature chiral spin liquids \cite{song2013mott} and magnetic orders such as generalized valence bond solids \cite{Corboz_2012_PRB}, plaquette \cite{Corboz_2013_PRB,Nataf_2016_PRB}, N\'{e}el and stripelike long-range \cite{Corboz_2011_PRL,Bauer_2012_PRB} or antiferromagnetic order \cite{weichselbaum2018unified}.

To observe these magnetic orders the atoms need to be cooled to temperatures below the superexchange energy $4t^2/U$, where $t$ and $U$ are the tunneling rate and interaction strength of the parent Hubbard model. Here, a promising approach is based on an (adiabatic) entropy redistribution akin to the Pomeranchuk effect in solid helium \cite{Richardson_1997_RMP}. For cold atoms in optical lattices this effect has been studied theoretically by means of dynamical mean field theory in \cite{Bernier_2009_PRA}, where the entropy was removed from a certain region - a dimple - by appropriately shaping the trapping potential. In the context of $\mathrm{SU}(N)$ fermions, Refs. \cite{Bonnes_2012_PRL,Hazzard_2012_PRA} have studied the enhancement of the cooling due to higher $N$ (see also \cite{Werner_2005_PRL} for adiabatic cooling of interacting and \cite{Blakie_2005_PRA,Blakie_2007_PRA} of non-interacting fermions).
Pomeranchuk and dimple cooling were experimentally demonstrated in \cite{Taie_2012_NatPhys} and \cite{Mazurenko2017} respectively, leading to an antiferromagnetic order \cite{Ozawa_2018_PRL,Chiu_2018_PRL} with \cite{Mazurenko2017} reporting the final temperature of $T/t = 0.25$ (see also \cite{Greif_2013_Science} for experimental realization of short-ranged antiferromagnetic order, \cite{PhysRevX.6.021030} for probing the Mott-insulator transition, and \cite{Sonderhouse2020} for the thermodynamics of the interacting $\mathrm{SU}(N)$ Fermi gas).

Motivated by these developments, in this work we study the effect of adiabatically loading an initally harmonically trapped $\mathrm{SU}(N)$ Fermi gas into a deep optical lattice in a species selective way: 
specifically, we consider a bi-partition of the atomic levels in two families, $A$ and $B$ such that $N = N_A + N_B$ and an optical potential which forms a dimple for only the $A$-family (hereafter we refer to the different atomic levels as colors). Using the high-temperature expansion of the Hubbard model we compute the entropy density and show that this results in further enhancement of the cooling of the Mott-insulating state of the $A$-family atoms in the dimple compared to a $\mathrm{SU}(N_A)$ Fermi gas only. 

The paper is structured as follows. In Sec. \ref{sec:Model} we describe the model and methodology, present the results in Sec. \ref{sec:Results}, discuss a possible experimental implementation in Sec. \ref{sec:Exp} and conclude in Sec. \ref{sec:Conclusions}.

\begin{figure}
  \centering
    \includegraphics{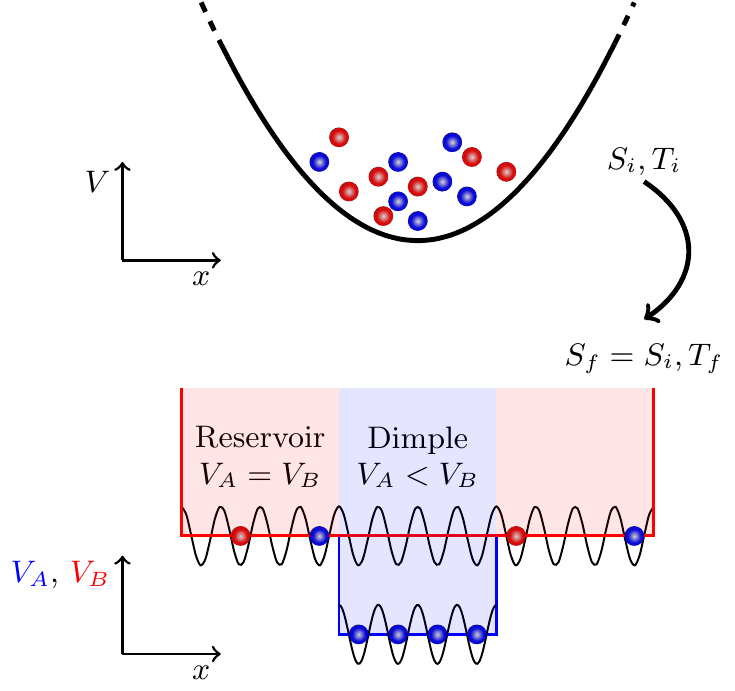}
	\caption{Schematics of the experimental protocol. A harmonically trapped $\mathrm{SU}(N)$ free Fermi gas with atoms belonging to families $A$ (blue) and $B$ (red) of initial total entropy $S_i$ and temperature $T_i$ is adiabatically loaded in a deep optical lattice with potentials $V_{A}$, $V_B$ for the two families such that $V_A < V_B$ in the dimple (blue shaded region) and $V_A = V_B$ in the reservoir (red shaded region).
}
	\label{fig:situationSketch}
\end{figure}

\section{The model}
\label{sec:Model}

Our main focus is to study the cooling of a $\mathrm{SU}(N)$ Fermi gas initially trapped in a harmonic potential. The trap is adiabatically transformed into a deep optical lattice, such that the system can be effectively described by a Hubbard model. We assume, that the final potential is such that a number $N_A$ out of the $N$ colors experience a different potential in a certain region of space - a dimple - than the remaining $N_B = N - N_A$ components, see Fig.~\ref{fig:situationSketch}.

Specifically, we consider a $\mathrm{SU}(N)$ Fermi gas of ${\cal N}_i = \sum_{\alpha=1}^N {\cal N}_\alpha$ particles, with ${\cal N}_\alpha$ the particle number of each color $\alpha$. We take the system to be initially a free gas in a harmonic potential $V({\bf r})=1/2 m \sum_{j=1}^d \omega_j^2 x_j^2$, where $m$ is the atom mass, ${\bf r}=(x_1,\ldots,x_d)$, $d$ the dimensionality of the system and $\omega_j$ the trapping frequencies with the geometric mean $\bar{\omega} = (\omega_1 \ldots \omega_d)^{1/d}$. Denoting further the chemical potential of each color as $\mu_\alpha$ and taking the gas to be at an initial temperature $T_i$, to first order in $T_i/\mu_\alpha$ the particle number and the chemical potential are related through (we use $\hbar = k_\mathrm{B} = 1$ throughout the article) \cite{Hazzard_2012_PRA}
\begin{equation}
\mathcal{N}_{i\alpha} = \frac{\mu_\alpha^d}{\bar{\omega}^d \, d!}.
\label{eq:Ni_sigma}
\end{equation}
The initial entropy of color $\alpha$ is then given by
\begin{equation}
S_{i\alpha} = T_i \, \frac{\mu_\alpha^{d-1}}{\bar{\omega}^d \, (d-1)!} \frac{\pi^2}{3}.
\label{eq:Si_sigma}
\end{equation}
Taking now into account the chemical potentials of each family, $\mu_A, \mu_B$, the total number of particles becomes
\begin{equation}
\mathcal{N}_{i} = \mathcal{N}_{iA} + \mathcal{N}_{iB},
\end{equation}
where
\begin{equation}
\mathcal{N}_{iF} = \sum_{\alpha \in F} {\cal N}_{i\alpha} = \frac{N_F \mu_F^d}{\bar{\omega}^d \, d!}
\end{equation}
is the particle number of family $F=A,B$, cf. the Eq.~(\ref{eq:Ni_sigma}).
Using that for a non-interacting gas the total initial entropy $S_i = \sum_\alpha S_{i\alpha}$, the entropy per particle is given by
\begin{equation}
\label{SperN_Ti_Harmonic}
\frac{S_{i}}{\mathcal{N}_{i}} = \frac{\pi^2}{3} d\frac{T_i}{T_{F,\text{eff}}}.
\end{equation}
Here,
\begin{equation}
T_{F,\text{eff}} = \frac{N_A \mu_A^d + N_B \mu_B^d}{N_A \mu_A^{d-1} + N_B \mu_B^{d-1}}
\label{eq:TFeff}
\end{equation}
is the effective Fermi temperature given by the weighted combination of the chemical potentials of both families.

Next, we assume that a deep optical lattice is loaded in an adiabatic, isentropic fashion, such that  the system is effectively described by a Hubbard Hamiltonian with tunneling rate $t$ and isotropic on-site interaction strength $U$ for all species \cite{Gorshkov_2010_NatPhys}
\beqa
\label{HubbardModel}
H &=& - t \sum \limits_{\braket{jk},\alpha} c^{\dagger}_{\alpha,j} c_{\alpha,k} + \sum \limits_{j,\alpha} V_{\alpha,j} \hat{n}_{\alpha,j} + \frac{U}{2} \sum \limits_j \hat{n}_j (\hat{n}_j - 1) \nonumber \\
&=& \sum_j h_j.
\eeqa
Here, $c_{\alpha,j}$ are the fermionic annihilation operators for a particle of color $\alpha$ on site $j$ with the usual anti-commutation relations $\{c_{\alpha,j},c^\dagger_{\beta,k}\}=\delta_{\alpha \beta} \delta_{j k}$, $\hat{n}_{\alpha,j} = c^{\dagger}_{\alpha,j} c_{\alpha,j}$ and $\hat{n}_j = \sum_\alpha  \hat{n}_{\alpha,j}$. The sum in (\ref{HubbardModel}) runs over $L$ sites, and $\langle jk \rangle$ denotes nearest neighbors.

A crucial ingredient of the present work are the species and position dependent on-site potentials $V_{\alpha,j}$. Here, we consider a different potential for each family, $V_{F,j} \equiv V_{\alpha,j}$ if $\alpha \in F, \; F=A,B$.
In particular, we consider box-like potentials, where $V_A < V_B$ in a central region which we call a \emph{dimple} ($D$). We denote the remainder of the sites as the \emph{reservoir} ($R$). Assuming box-like potentials is motivated by the fact, that in a quantum simulation of $\mathrm{SU}(N)$ magnetism, one ideally wishes to create a flat optical lattice to faithfully simulate the Hubbard model. There is indeed an ongoing effort to achieve this goal in current cold-atom experiments \cite{Mazurenko2017} as well as in creating box-shaped rather than harmonic potentials \cite{Gaunt_2013_PRL}. Without loss of generality we choose the potentials as
\begin{subequations}
	\label{eq:V}
	\begin{align}
	V_{A,j} &= 
	\begin{cases}
		0 \; \phantom{V_A} {\rm for} \; j \in R  \\
		V_A \; \phantom{0} {\rm for} \; j \in D
    \end{cases} \label{eq:VA}
	\\
    V_{B,j} &= 0 \; \; \forall j \label{eq:VB}
    \end{align}
\end{subequations}
with $V_A < 0$, see Fig. \ref{fig:situationSketch}.
In what follows, we analyze the two-family Hubbard model using its high-temperature expansion in the grand canonical setting \cite{oitmaa} and local density approximation (LDA), which is commonly adopted for deep optical lattices realizing the tight-binding models \cite{Taie_2012_NatPhys, Hazzard_2012_PRA, Ozawa_2018_PRL} (we further comment on the applicability of LDA for the box potentials below). The particle and entropy densities at site $j$ are given by ($F=A,B$)
\beqa
	\bar{n}_{F,j} &=& - \partial_{\mu_F} \Omega_j \label{eq:ni} \\
	s_j &=& - \partial_T \Omega_j, \label{eq:si}
\eeqa
where $\Omega_j$ is the local contribution to the grand potential, cf. Eq.~(\ref{eq:Omega}). Furthermore we define the entropy density \emph{per particle} as 
\beq
	\bar{s}_j = \frac{s_j}{\bar{n}_{A,j} + \bar{n}_{B,j}}.
\eeq
~\\

\begin{figure}
  \centering
	\includegraphics[width=0.5\textwidth]{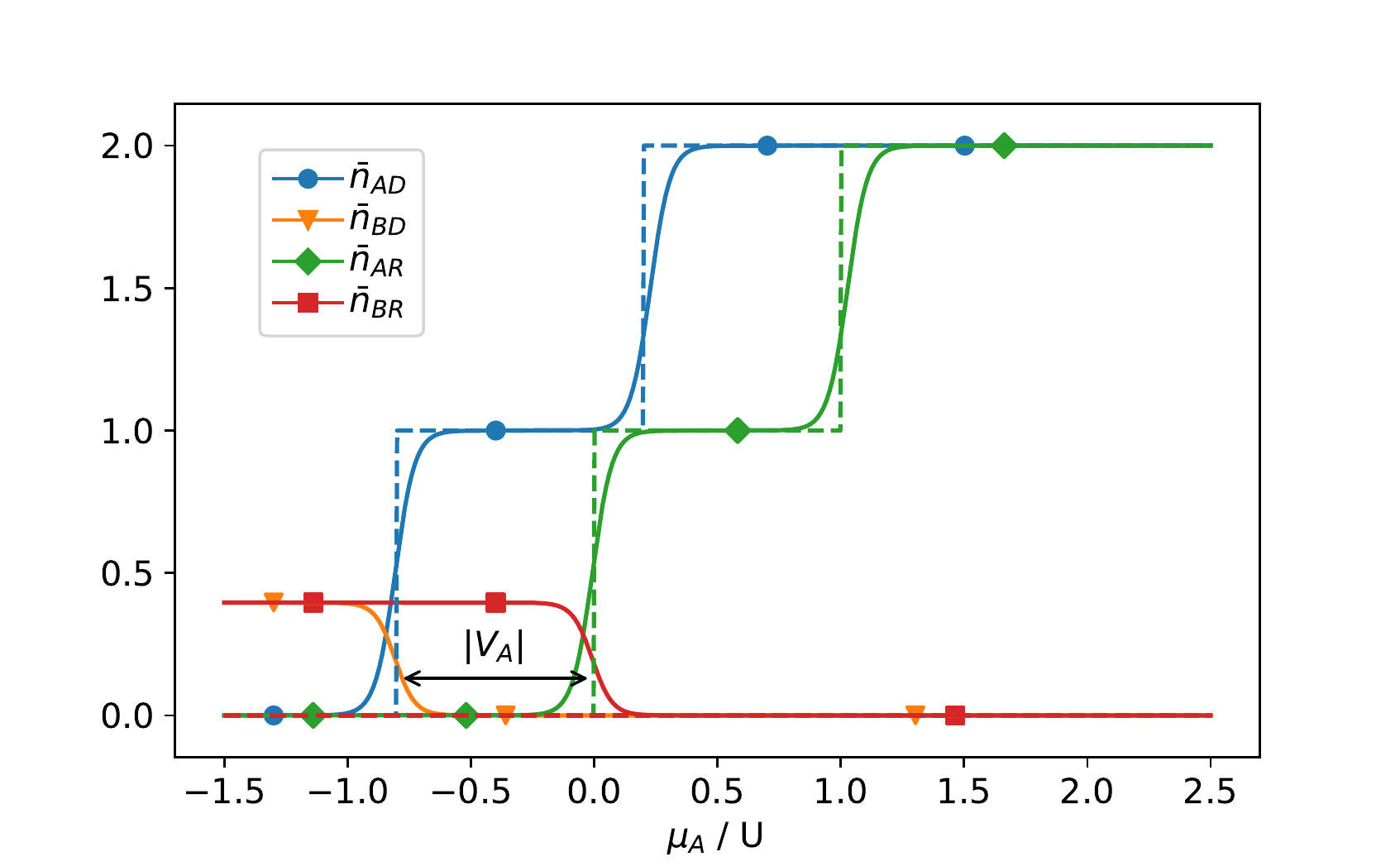}
	\caption{
	Single-site particle densities in the dimple $\bar{n}_{AD}, \, \bar{n}_{BD}$ and in the reservoir $\bar{n}_{AR}, \, \bar{n}_{BR}$ vs. $\mu_A$. The densities are plotted in the atomic limit, $t=0$, for $T=0$ (dashed lines) and $T = U / 25$ (solid lines), $N_A = 2, \, N_B = 8$, $V_A = -0.8\,U$, cf. Eq.~$\eqref{eq:V}$, and at fixed $\mu_B = -0.1U$ (the densities $\bar{n}_{BD}$, $\bar{n}_{BR}$ are identically zero at zero temperature). The arrow depicts the offset $|V_A|$ between the dimple and the reservoir particle densities, see text for details.
}
	\label{fig:ChemicalPotentialParticleDensities}
\end{figure}

\textit{The atomic limit.}
We start our analysis by first considering a single site in the atomic limit $t=0$. The single site partition function is given by $z_{0,j} = \textrm{tr} \, e^{-\beta h_j}$, where $h_j$ is a single site Hamiltonian in Eq. (\ref{HubbardModel}) and the trace is taken over a basis of single-site orbitals of $h_j$. In this case, the single site partition function reads
\begin{equation}
\label{TwoVarietySum}
z_{0,j} = \sum \limits_{n_A = 0}^{N_A} \sum \limits_{n_B = 0}^{N_B} {N_A \choose n_A} {N_B \choose n_B} e^{-\beta \epsilon_j(n_A,n_B)},
\end{equation}
where $\beta = 1/ T$ and 
\begin{equation}
\label{EnergyFunction}
\begin{split}
\epsilon_j(n_A,n_B) = &(n_A + n_B) (n_A + n_B - 1) U / 2 \\
+ (V_{A,j} - & \mu_A) \, n_A + (V_{B,j}- \mu_B) \, n_B.
\end{split}
\end{equation}

It is instructive to consider further the limit of small temperatures and investigate the behavior of the particle densities (\ref{eq:ni}) in the dimple and the reservoir as functions of the chemical potentials $\mu_F$. For $\beta \gg 1$, the partition function (\ref{TwoVarietySum}) is dominated by a single term, corresponding to the minimum of the energy (\ref{EnergyFunction}), with the particular combination of $(n_A,n_B)$ such that $n_A=\bar{n}_A, n_B=\bar{n}_B$ and (\ref{TwoVarietySum}) reduces to
\begin{equation*}
z_{0,j} \approx {N_A \choose \bar{n}_{A}} {N_B \choose \bar{n}_{B}} e^{-\beta \epsilon_j(\bar{n}_A,\bar{n}_B)}.
\end{equation*}
Consequently, the entropy density is given by
\begin{equation}
s_j = \text{log}\left({N_A \choose \bar{n}_{A}} {N_B \choose \bar{n}_{B}} \right).
\end{equation}
For specificity, in what follows we seek to create a ``clean'' Mott-insulating state with $\bar{n}_A = 1$ and no B-particles, $\bar{n}_B=0$, in the dimple, a scenario we analyze in detail in Sec.~\ref{sec:Results}.
In this case, the value of $V_A$ has to be chosen in the interval $(-U,0)$ avoiding the proximity of the limiting values $V_A=-U,0$. This is to prevent possible double occupancies (when $V_A = -U$) and to ensure $\bar{n}_A=1$ (avoiding too shallow dimple $V_A = -\epsilon, \; \epsilon \ll 1$) at finite temperature. We have found that these constraints are well respected for $V_A = -0.8U$ which we consider in the remainder of the paper.
We also note that $\bar{n}_{AR} < \bar{n}_{AD}$ as a consequence of he dimple potential Eq.~\ref{eq:V}.

Analogously, as discussed in detail in Appendix~\ref{app:Densities}, a suitable choice of the chemical potential for the B-family is $\mu_B < 0$ in which case $\bar{n}_{BD} = \bar{n}_{BR} = 0$ at zero temperature and $\bar{n}_A$ undergoes changes in integer steps ($0 \rightarrow 1 \rightarrow \ldots \rightarrow N_A$) as $\mu_A$ is increased from $-\infty$ to positive values, cf. the dashed lines in Fig.~\ref{fig:ChemicalPotentialParticleDensities}. The transitions from $\bar{n}_A$ to $\bar{n}_A+1$ occur at $\mu_A = V_A + \bar{n}_A U$ in the dimple and $\mu_A = \bar{n}_A U$ in the reservoir, which differ by $V_A$, as indicated by the arrow in Fig.~\ref{fig:ChemicalPotentialParticleDensities}. 

The effect of the finite temperature is the characteristic ``smearing'' of the staircase profile of the particle densities as well as resulting in $
\bar{n}_{B} > 0$ in the $\mu_A \rightarrow -\infty$ limit, cf. the orange and red solid lines in Fig.~\ref{fig:ChemicalPotentialParticleDensities}. The precise value of $\bar{n}_{AR}, \bar{n}_{BR}$ can be further adjusted by $\mu_{A,B}$, which we tune in the vicinity of 0, cf. Fig.~\ref{fig:ChemicalPotentialParticleDensities}, such that the Mott-insulating state is achieved in the dimple, cf. Sec.~\ref{sec:Results} and Appendix~\ref{app:Densities} for further details.

~\\
\textit{The $t/U$ expansion at finite temperature.}
Next, we turn to the $t \neq 0$ regime. Since we assume a box-shaped potential, the LDA is satisfied everywhere but at the boundary between the dimple and the reservoir, where the potential $V_A$ changes in a step-like fashion. For large enough reservoir and dimple, we expect the thermodynamic properties of the Fermi gas far from the boundary between the two regions to be still well captured by the LDA. Under this approximation, the grand-canonical potential of the two-family Hubbard model Eq.~(\ref{HubbardModel}), up to second order in $t/U$ for $t \ll T \ll U$, reads \cite{oitmaa}
\begin{equation}
	\label{eq:Omega}
	\Omega = \sum_{j=1}^L \Omega_j = -\beta^{-1} \sum_{j=1}^L \text{log}(z_{0,j}) + \sum_{j=1}^{L} \Omega_{2,j},
\end{equation}
where $L=L_D + L_R$, $L_{D,R}$ being the number of sites in the dimple and the reservoir respectively and (see Appendix \ref{app:SecondOrder} for derivation)
\begin{widetext}
\begin{equation}
\label{TwoVarietySecondOrderTerm}
\begin{split}
\Omega_{2,j} = & -\beta^{-1} t^2 c_\ell  z_{0,j}^{-2} \, \sum \limits_{F = A,B} \Bigg[ N_F \sum \limits_{n_{1F} = 1}^{N_F} \sum \limits_{n_{1 \bar{F}} = 0}^{N_{\bar{F}}}  \sum \limits_{n_{2F} = 0}^{N_F - 1} \sum \limits_{n_{2{\bar{F}}} = 0}^{N_{\bar{F}}} \times \\
& e^{-\beta (\epsilon_j(n_{1A},n_{1B}) + \epsilon_j(n_{2A},n_{2B}))} {N_F - 1 \choose n_{1F} - 1} {N_F - 1 \choose n_{2F}} {N_{\bar{F}} \choose n_{1{\bar{F}}}} {N_{\bar{F}} \choose n_{2{\bar{F}}}} I(U(n_{1A} + n_{1B} - n_{2A} - n_{2B} - 1)). \Bigg]
\end{split}
\end{equation}
\end{widetext}
Here, $\bar{F}$ denotes the complement of the family $F$, i.e. either $F=A, \bar{F}=B$ or vice versa,
$c_\ell$ is the coordination number of the lattice, the energies $\epsilon_j(n_{A},n_{B})$ are given by \eqref{EnergyFunction} and the function $I$ is given by
\begin{equation}
\label{theIntegralFunction}
I(\Delta) = 
\begin{cases}
\frac{\beta^2}{2} & , \Delta = 0\\
\frac{1}{\Delta^2}(e^{\beta \Delta} - \beta \Delta - 1) & , \Delta \neq 0.
\end{cases}
\end{equation}

\section{Results}
\label{sec:Results}
\begin{figure}
  \centering
	\includegraphics[width=0.5\textwidth]{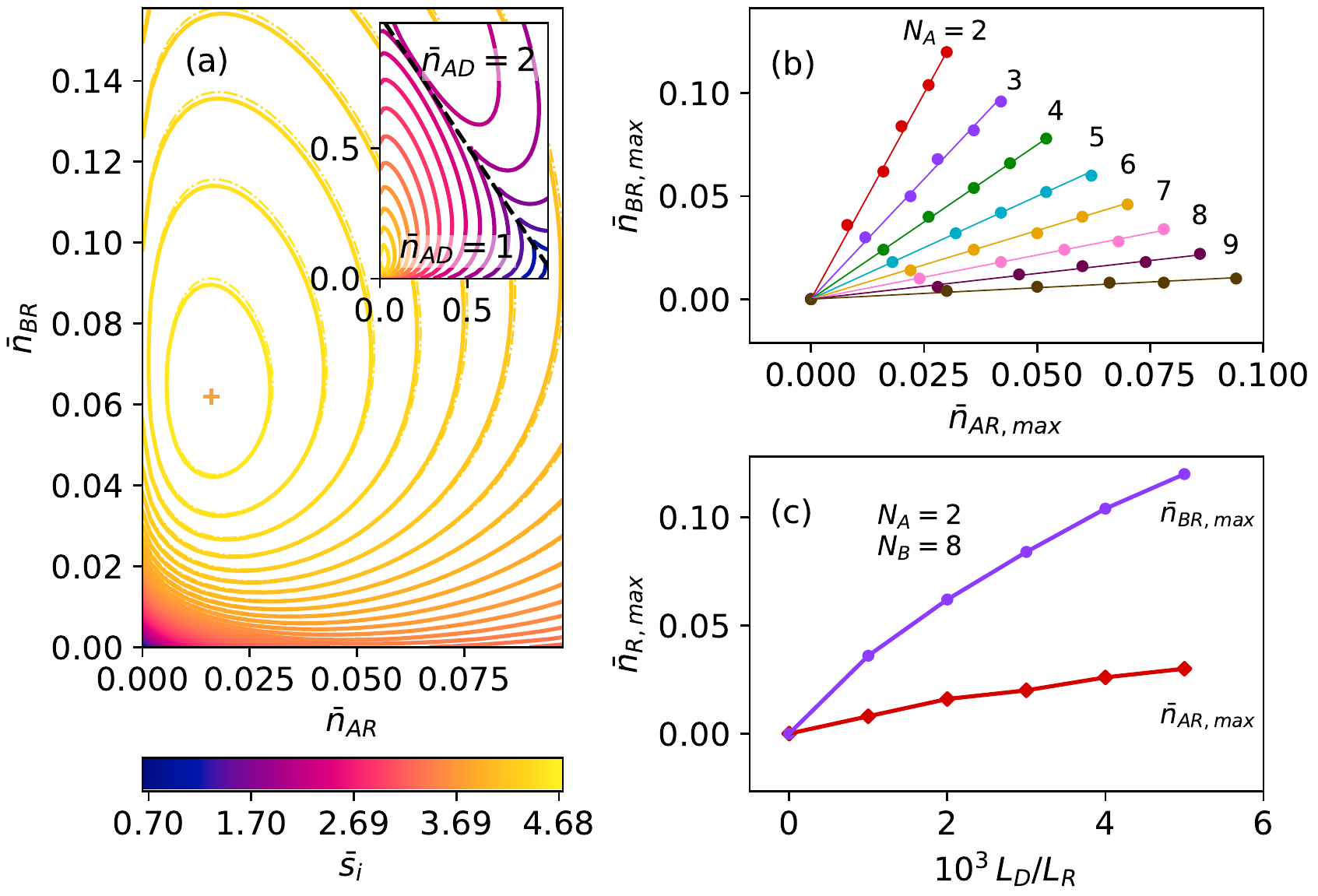}
	\caption{
	{\bf (a)} Isolines of the initial entropy density per particle $\bar{s}_i$ as a function of the particle densities in the reservoir at fixed $T_f=4t$. The cross indicates the location $(\bar{n}_{AR,{\max}},\bar{n}_{BR,{\max}})$ of maximum  of $\bar{s}_i$. The dashed (solid) lines correspond to the atomic limit (second order high-temperature expansion) of the Hubbard model respectively. The inset shows a larger range of reservoir particle densities, with a black dashed line delimiting the Mott-insulating regions $\bar{n}_{AD}=1,2$ in the dimple.
	{\bf (b)} $\bar{n}_{AR,{\rm max}}$ vs. $\bar{n}_{BR,{\rm max}}$ for various $N_A$. The data points correspond to various dimple/reservoir sizes $L_D/L_R$ indicated in pane (b).
	{\bf (c)} $\bar{n}_{AR,{\rm max}}$ and $\bar{n}_{BR,{\rm max}}$ as a function of the relative size of the dimple and the reservoir $L_D/L_R$.
	Parameters used: $U/t=100$, $V_A = -0.8 \, U$, and $L_D/L_R = 1/50$. In (a,c): $N_A=2,N_B=8$.
}
	\label{fig:entropy_isolines}
\end{figure}
For the present simulations, we consider a two-dimensional square lattice with coordination number $c_\ell=4$. Motivated by possible applications in ongoing experiments with ${}^{87}{\rm Sr}$ atoms, we also fix $N=10$  \cite{DeSalvo_2010_PRL, Stellmer_2013_PRA, stellmer2014degenerate}.

~\\
\textit{Particle densities in the dimple and the reservoir.}
We start our investigations by discussing the role of the particle densities. 
It follows from the form of the potential for family $A$, Eq.~(\ref{eq:VA}), and the discussion in Sec. \ref{sec:Model}, that as $\mu_A$ is increased, particles of family $A$ will accumulate in the dimple until they reach unit filling. Upon further increase of $\mu_A$, they will start to populate the reservoir, see Fig. \ref{fig:ChemicalPotentialParticleDensities}. Subsequently, when increasing $\mu_B$, for $\mu_B < U$, particles of family $B$ will start to populate only the reservoir as they will be repelled from the dimple by particles $A$ present therein. 
Focusing specifically on the range of chemical potentials resulting in $\bar{n}_{A,D}=1$ (cf. the inset of Fig.~\ref{fig:entropy_isolines}a), in Fig.~\ref{fig:entropy_isolines}a we show the dependence of the entropy density per particle $\bar{s}_i = S_i/({\cal N_A} + {\cal N}_B) = (L_R s_R + L_D s_D)/({\cal N_A} + {\cal N}_B)$ at a given final temperature ($T_f=4t$) as a function of the particle densities. 
Ultimately, we seek conditions which minimize the entropy density per particle $\bar{s}_D$ in the dimple, which we analyze in the subsequent section. Alternatively, one can invert the question and ask, given the final temperature $T_f$, what parameter set maximizes the (total) initial entropy density per particle $\bar{s}_i$.
It is apparent from Fig. \ref{fig:entropy_isolines}a, that there is a unique combination of the particle densities $\bar{n}_{AR,{\rm max}}, \bar{n}_{BR,{\rm max}}$, denoted by a cross, which maximizes $\bar{s}_i$. Two comments are in order -- first, the fact that $n_{B,\rm max}>0$ clearly indicates an improved cooling due to the presence of the family $B$. Intuitively, this is an expected result, since the presence of family $B$ increases the number of degrees of freedom in the reservoir which are able to absorb the entropy from the dimple. 
Second, starting from the partition function in the atomic limit (\ref{TwoVarietySum}) in the regime $\bar{n}_A,\bar{n}_B, < 1$, in the Appendix \ref{app:Symmetry} we show that $(\bar{n}_{AR,{\rm max}}, \bar{n}_{BR,{\rm max}})$ corresponds to the symmetric point $\mu_A = \mu_B$ restoring the $\mathrm{SU}(N)$ Hubbard model in the reservoir.
In Fig. \ref{fig:entropy_isolines}b we show the dependence of $\bar{n}_{BR,{\rm max}}$ on $\bar{n}_{AR, {\rm max}}$ for various $N_A$ and $L_D / L_R$ denoted by the data points in Fig.~\ref{fig:entropy_isolines}c. This dependence can be understood by considering the atomic limit, in which $N_B \, n_{BR,{\rm max}} = N_A \, n_{AR, {\rm max}}$, which follows directly from the properties of the partition function (\ref{TwoVarietySum}) [see Appendix \ref{app:Symmetry}].

Next, in Fig.~\ref{fig:entropy_isolines}c we show the dependence of $\bar{n}_{FR,{\rm max}}$ vs. $L_D/L_R$. This is motivated by the requirement that within the finite amount of space available to the experiment, one has a trade off between the size of the dimple and the reservoir. In order to optimize the cooling, one has to adjust the particle densities in the reservoir. In particular, in the limit of infinite reservoir size $L_D/L_R \rightarrow 0$ the optimal cooling is achieved for $\bar{n}_{FR,{\rm max}} \rightarrow 0$
\footnote{In this context Ref. \cite{Bernier_2009_PRA} discusses the improvement in cooling when flattening the harmonic profile of the reservoir, resulting in flat (box-like) profile considered here.}.

We now turn our attention to the cooling in the dimple, where we compare the cooling in the presence of family $B$ with the situation when it is absent, the latter corresponding to the $\mathrm{SU}(N_A)$ Hubbard model only.

%
%
%
\begin{figure}
  \centering
	\includegraphics[width=0.5\textwidth]{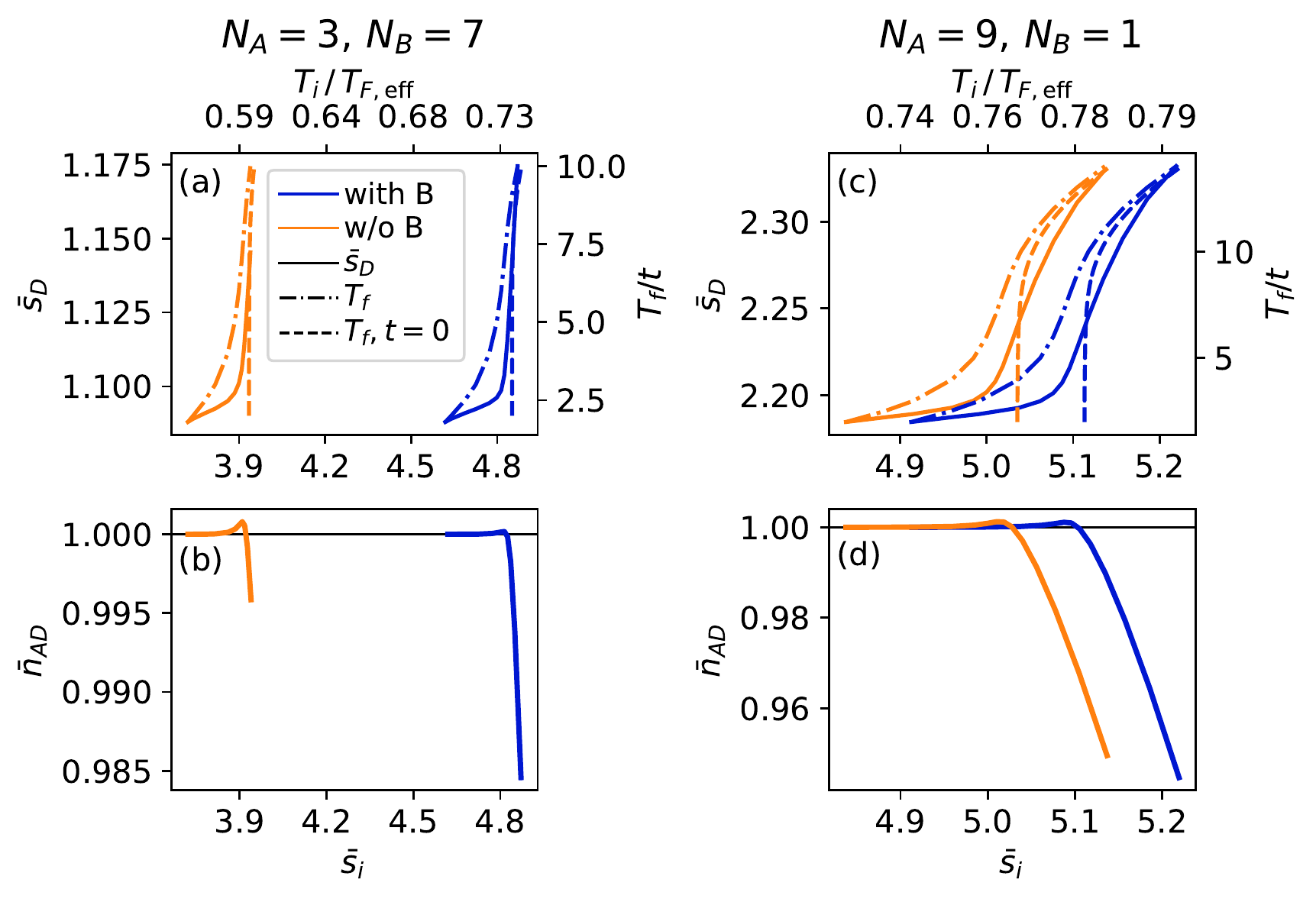}
	\caption{
	Entropy density per particle in the dimple $\bar{s}_D$ vs. the initial entropy density $\bar{s}_i$ for (a) $N_A=3$ and (c) $N_A=9$.	
	The dark blue (light orange) curves correspond to situations with (without) family B. The solid, dash-dotted and dashed lines correspond to the entropy density $\bar{s}_D$ and the final temperature $T_f$ to second order expansion Eq.~(\ref{TwoVarietySecondOrderTerm}) and in the atomic limit respectively. 
	On the top horizontal axis of panels (a) and (c) we show $T_i / T_{F,\text{eff}}$, where $T_{F,\text{eff}}$ is the effective Fermi temperature Eq.~(\ref{eq:TFeff}). Panels (b),(d) show the corresponding particle densities in the dimple $\bar{n}_{AD}$.
	}
	\label{fig:sVariation}
\end{figure}
~\\
\textit{Dimple cooling.}
Using the analysis described above, for each $T_f$ we find maximum $\bar{s}_i$ and evaluate the entropy density per particle in the dimple $\bar{s}_D$. The dependence of $\bar{s}_D$ and $T_f$ on $\bar{s}_i$ is shown in Fig. \ref{fig:sVariation}a and Fig. \ref{fig:sVariation}c for $N_A=3$ and $N_A=9$ respectively. Figs. \ref{fig:sVariation}b,d show the corresponding particle densities in the dimple. For illustration we also show the corresponding initial temperatures $T_i$ evaluated using Eq.~(\ref{SperN_Ti_Harmonic}) and specific experimental parameters, see caption for details. It is apparent from the figures that the improvement in cooling, i.e. achieving the same $\bar{s}_D$ for a larger initial entropy density, increases with increasing $N_B$. We further note that the atomic limit predictions (dashed lines in Figs. \ref{fig:sVariation}a,c) saturate for a certain $\bar{s}_i$ at $\bar{s}_D = \log \, N_A$ signaling the necessity to include higher order terms Eq. (\ref{TwoVarietySecondOrderTerm}) to capture the behavior of the entropy in the dimple. The relatively small change in $\bar{s}_D$ can be attributed to the fact that for high temperatures $T_f \gg t$ considered here the entropy density is only weakly dependent on the temperature
\footnote{See e.g. \cite{Bonnes_2012_PRL} or Fig. 1 in \cite{Messio_2012_PRL}, which analyzed the entropy density for a one-dimensional chain with $c_\ell=2$. Since we rely on LDA, we expect the dependence of $\bar{s}_D$ to qualitatively hold for the square lattice with $c_\ell=4$ as it appears only as a prefactor in Eq.~(\ref{TwoVarietySecondOrderTerm}).}. 

Addressing quantitatively the regime of small final temperatures $T_f \lesssim t$ relevant for the superexchange physics would require different theoretical tools, such as dynamical mean-field theory (DMFT) \cite{Bernier_2009_PRA,Jordens_2010_PRL} or Quantum Monte Carlo or tensor network based approaches \cite{PhysRevX.5.041041}. 
The complexity of adapting these methods to the problem of the two-family ${\rm SU}(N)$ Hubbard model goes beyond the scope of this work. 
However, in Appendix~\ref{app:DMFT} we compare the employed (second-order) high-temperature expansion to the DMFT results of Ref.~\cite{Bernier_2009_PRA} for a ${\rm SU}(2)$ Hubbard model with a three-dimensional dimple. We find a good agreement, similarly to Ref.~\cite{Jordens_2010_PRL}, between the two methods in the expected regime of validity $T_f \gtrsim t$. 
This agreement is a strong indication in favor of the quantitative correctness of the data shown in Fig.~\ref{fig:sVariation}, which clearly indicate the enhancement of the cooling when considering the $B$-family in the reservoir as compared to the case when no $B$-family is present.

\section{Experimental considerations}
\label{sec:Exp}

In this section we briefly discuss a possible implementation of the proposed scheme. 
We seek parameters that satisfy the following constraints: (\emph{i}) a deep optical lattice with potential amplitude $V_{\rm latt} \approx O(10 E_r)$, where $E_r = (\hbar k_{\rm latt})^2/(2m)$ is the recoil energy, such that the tight binding approximation holds, (\emph{ii}) the lattice band gap, which for the deep lattice we estimate as a single lattice site harmonic oscillator frequency $E_{\rm gap} \approx \sqrt{2 V_{\rm latt} k_{\rm latt}/m}$, to be much larger than the interaction energy to neglect higher band excitations, $E_{\rm gap} \gg U$, and (\emph{iii}) a negligible off-resonant scattering rate with respect to the Hamiltonian energy scales. For the sake of concreteness, in the following we specifically focus on fermionic ${}^{87}{\rm Sr}$ \cite{DeSalvo_2010_PRL, Stellmer_2011_PRA, Stellmer_2013_PRA, stellmer2014degenerate} and provide a quantitative example restoring the dimensionful quantities using $\hbar$.

In the far-detuned regime, the optical potential and off-resonant scattering rate are given by the classical formulas $V = -3\pi c^2/(2 \omega_0^3) \gamma [1/(\omega_0 - \omega) + 1/(\omega_0 + \omega)] I$ and $\gamma_{\rm sc} = 3\pi c^2/(2 \hbar \omega_0^3) (\omega/\omega_0)^3 \gamma^2 [1/(\omega_0 - \omega) + 1/(\omega_0 - \omega)]^2 I$, where $\omega_0$, $\omega$, $\gamma$ and $I$ are the atomic transition frequency, the laser light frequency, the atomic excited state decay rate and the laser intensity respectively \cite{grimm2000optical}.

We shall consider the dimple potential to be created by a laser light on the $\ket{\mathbb S} - \ket{\mathbb P}$ transitions, where $\ket{\mathbb S} \equiv \ket{{}^1{\rm S}_0, F=9/2}, \ket{\mathbb P} \equiv \ket{{}^3{\rm P}_2, F'=11/2}$ for brevity \cite{Onishchenko_2019_PRA}. 
The choice of the ${\mathbb P}$ manifold is motivated by the fact that the main optical lattice wavelength $\lambda_{\rm latt} = 900\;{\rm nm}$ is approximately magic for the $\ket{\mathbb S}-\ket{{\mathbb P}}$ transition
\footnote{Alex Urech, private communication}
which ensures a position independent frequency selection of the individual $m_{F}$ states. To this end, a laser intensity of the lattice $I_{\rm latt} = 5 \, {\rm k W/cm}^2$ yields $V_{\rm latt}/E_r \approx 20$ and with $U = 5 \, {\rm kHz}$ we get $E_{\rm gap} \approx 160 \, {\rm kHz} \gg U$ as desired. We also anticipate that the dominant scattering rate corresponds to the scattering of the lattice light on the $\ket{\mathbb S} - \ket{{}^1 {\rm P}_1}$ transition and evaluates to $\gamma_{\rm sc} \approx 6 \, {\rm mHz}$, which is negligible compared to the Hamiltonian energy scales.

Next, denoting $\Delta = \omega_0 - \omega$ and requiring that $|\Delta| \gg |V_A|$ such that the far-detuning approximation holds, we find that the desired $V_A \approx -U$ is achieved for $I \approx 20 \, {\rm W/cm}^2$ and $\Delta = 50 \, {\rm kHz}$. This value of $\Delta$ is compatible with the single $m_F$-level addressability using the Zeeman splitting of the ${\mathbb P}$-manifold with the
energy shift between adjacent $m_F$ states of $0.255\,{\rm MHz}/{\rm G}$ giving, say,  $25 \, {\rm MHz}$ for a magnetic field of $100 \, {\rm G}$ \cite{Boyd_2007_PRA}, cf. also \cite{Taie_2010_PRL} for experimental demonstration using ${}^{173}{\rm Yb}$.

Importantly, the dimple light gives rise to additional contribution to the dimple potential $\delta V \approx N_A \times 2 \, {\rm kHz}$ for \emph{all} $m_F$ states stemming from the $\ket{\mathbb S} - \ket{{}^1{\rm P}_1}$ transition, which is of the order comparable to the target dimple offset $U$. Here the factor $N_A$ accounts for the $N_A$ dimple laser beams. 
In principle one could mitigate this additional potential by reducing further $\Delta$ (while modifying the dimple laser intensity $I$ to keep $|V_A| \approx U$), however this is precluded by the requirement $|\Delta| \gg |V_A|$ so that one remains in the far-detuned regime to prevent detrimental light scattering.
A possible remedy is to compensate for the additional dimple potential $\delta V$ with 
a dipole laser beam in the dimple that is blue detuned to the $\ket{\mathbb S} - \ket{{}^1 {\rm P}_1}$ transition or, alternatively, a red-detuned one in the reservoir region.

Finally, we note that using ${}^{173}{\rm Yb}$ instead might provide further improvement in reducing the additional dimple potential \cite{Berends_1992_JOSA,Shibata_2014_PRA,Boyd_2007_PRA,Ludlow_2015_RMP,Takasu_2017_PRA}. This stems from the stronger $\ket{\mathbb S}-\ket{\mathbb P}$ transition
with the decay rate of $\approx 6 \, {\rm mHz}$ for ${}^{87}{\rm Sr}$ and $\approx 95 \, {\rm mHz}$ for ${}^{173}{\rm Yb}$. This in turn allows for a reduction of the dimple laser intensities and consequently of the additional dimple potential by a factor of $95/6 \approx 15$.

\section{Conclusions and Outlook}
\label{sec:Conclusions}

We have studied the enhancement of cooling of a $\mathrm{SU}(N)$ Fermi gas exploiting state selective trapping of a subset of $N_A$ atomic levels for which the trapping potential forms a dimple. 
We could demonstrate such enhancement and quantify the cooling using the high-temperature expansion of the Hubbard model by explicit evaluation of the entropy densities and final temperatures leading to a $\mathrm{SU}(N_A)$ Mott-insulator in the dimple. 
We could also demonstrate that optimal cooling occurs when the chemical potentials for both families are equal in the reservoir, leading to the symmetry restoration of the $\mathrm{SU}(N)$ Hubbard model therein. 
While these results are encouraging for the current experiments with cold fermionic gases featuring $N$ sub-levels, such as ${}^{173}$Yb or ${}^{87}$Sr, the high-temperature expansion used here is not suitable to describe the regime of sufficiently small temperatures where exotic magnetic phases driven by the superexchange interaction could be achieved.
To faithfully quantify the cooling at such low final temperatures $T_f < t$ requires implementing some of the methods discussed in Sec.~\ref{sec:Results}, such as the DMFT \cite{Bernier_2009_PRA,Jordens_2010_PRL} or some of the Quantum Monte Carlo or tensor-network based approaches \cite{PhysRevX.5.041041}, which we leave for future work.

\section{Acknowledgments}
We are very grateful to Kilian Sandolzer, Tilman Esslinger, Tobias G\"unther, Alex Urech, Benjamin Pasquiou and Kaden Hazzard for useful discussions. This work has received funding from the European Research Council (ERC) under the European Union's Seventh Framework Programme (FP7/2007-2013) (Grant agreement No. 615117, QuantStro) and
the Netherlands Organisation for Scientific Research (NWO) (Grant No. 024.003.037, Quantum Software Consortium).




\input{SUN_Cooling.bbl}

\appendix

\section{Particle densities in the atomic and zero temperature limit}
\label{app:Densities}

Here we discuss the particle densities in the dimple and the reservoir in the atomic and zero temperature limit. The particle densities are given by Eq.~(\ref{eq:ni}), which in the atomic limit and using LDA reduces to
\beq
	\bar{n}_{F,j} = -\partial_{\mu_F} \Omega_{0,j} = \frac{\sum \limits_{n_A = 0}^{N_A} \sum \limits_{n_B = 0}^{N_B} {N_A \choose n_A} {N_B \choose n_B} e^{-\beta \epsilon_j(n_A,n_B)} n_F }{z_{0,j}},
	\label{eq:n_app}
\eeq
where we have used the expression $\Omega_{0,j} = -1/\beta \log z_{0,j}$ for the atomic limit grand potential, cf. Eq.~(\ref{eq:Omega}), and $F=A,B$.
We note that in the infinite temperature limit $\beta \rightarrow 0$ the expression for the particle densities Eq.~(\ref{eq:n_app}) reduces to $\bar{n}_{F,j} = N_F/2$ which is the expected result as all the particle numbers become equally likely. On the other hand, in the zero temperature limit $\beta \rightarrow \infty$, the Eq.~(\ref{eq:n_app}) is dominated by a single term with the \emph{lowest} energy $\epsilon_j$, cf. Eq.~(\ref{EnergyFunction}), which we rewrite as (dropping the site index $j$ for simplicity and setting $\tilde{V}_B=0$, cf. the Eq. (\ref{eq:V}))
\beq
	2 \tilde{\epsilon} = n_A^2 + n_B^2 + 2 n_A n_B + n_A(2 \tilde{V}_A - 2 \tilde{\mu}_A - 1) - n_B (2\tilde{\mu}_B + 1).
	\label{eq:energy}
\eeq
Here we have denoted by tilde the quantities rescaled by the interaction energy, $\tilde{\epsilon} = \epsilon/U$, $\tilde{V}_F = V_F/U$, $\tilde{\mu}_F = \mu_F/U$. It should be noted that the fact that the sum in Eq.~(\ref{eq:n_app}) is dominated by a single term of given $n_A, n_B$ implies that the \emph{particle densities} correspond to these, $\bar{n}_F = n_F$. In order to determine the particle numbers $\bar{n}_F$ as a function of $\tilde{\mu}_F$ it thus suffices to identify the combination $(n_A,n_B)$ which minimizes the energy (\ref{eq:energy}) for a given set of parameters $\tilde{\mu}_F$, $\tilde{V}_A$. 

To demonstrate this, let us first consider a limit $\tilde{\mu}_A \rightarrow -\infty$, such that the lowest energy corresponds to $n_A = 0$ and Eq.~(\ref{eq:energy}) becomes
\beq
	2 \tilde{\epsilon}(n_A=0,n_B) = n_B \left( n_B - 1 - 2 \tilde{\mu}_B \right).
	\label{eq:energyB}
\eeq
Similarly, the minimum of (\ref{eq:energyB}) implies $n_B=0$ for $\tilde{\mu}_B \rightarrow -\infty$. Increasing $\tilde{\mu}_B$
then leads to a series of transitions, in steps of 1, in the particle number $\bar{n}_B$ and the threshold values of $\tilde{\mu}_B$ can be obtained from the relation
\beq
	\tilde{\epsilon}(0,n_B) = \tilde{\epsilon}(0,n_B+1)
\eeq
which leads to
\beq
	\tilde{\mu}_B^{(n_B \, \leftrightarrow \, n_B + 1)} = n_B.
	\label{eq:muB}
\eeq
This allows us to analyze the situation of Fig.~\ref{fig:ChemicalPotentialParticleDensities} and to identify the particle numbers as $\tilde{\mu}_A$ is varied. For $\tilde{\mu}_B = -0.1$ the Eq.~(\ref{eq:muB}) implies $n_B=0$. As we increase $\tilde{\mu}_A$ from $-\infty$, more A-particles will populate the dimple and the reservoir and thus $n_B$ remains zero. The energy (\ref{eq:energy}) simplifies to
\beq
	2 \tilde{\epsilon}(n_A,n_B=0) = n_A \left(n_A + 2 \tilde{V}_A - 2 \tilde{\mu}_A - 1 \right).
\eeq
From the condition $\tilde{\epsilon}(n_A,0) = \tilde{\epsilon}(n_A+1,0)$ we get the threshold values for $\tilde{\mu}_A$
\beq
	\tilde{\mu}_A^{(n_A \leftrightarrow n_A+1)} = \tilde{V}_A + n_A.
\eeq
for which the number of A-particles changes from $n_A$ to $n_A+1$ until the saturation $n_A = N_A$ for $\tilde{\mu}_A > \tilde{V}_A + N_A - 1$. 

In principle it is straightforward to extend this analysis to other set of parameters, which we do not perform explicitly as we are mainly interested in the parameter regime of vanishing density of B-particles in the reservoir.

~\\
\textit{Finite temperature.} The effect of finite temperature is to ``smear'' out the staircase structure of $\bar{n}_A$ as is apparent from the Fig.~\ref{fig:ChemicalPotentialParticleDensities}. Similarly, we note that for the parameters of Fig.~\ref{fig:ChemicalPotentialParticleDensities} the non-zero value of $\bar{n}_B$ in the $\tilde{\mu}_A \rightarrow -\infty$ limit is the consequence of non-zero temperature, which interpolates between $\bar{n}_B=0$ for $\beta \rightarrow \infty$ and $\bar{n}_B = N_B/2$ for $\beta = 0$.

\section{Derivation of the Eq.~(\ref{TwoVarietySecondOrderTerm})}
\label{app:SecondOrder}

In this section we provide the details of the derivation of the Eq.~(\ref{TwoVarietySecondOrderTerm}) following closely the treatment in \cite{Henderson_1992_PRB} and \cite[chap. 1,7,8]{oitmaa} (cf. also \cite{Pan_1991_PRB, pan1991linked, thompson1991high} for related developments). 
It is obtained using the high-temperature expansion of the Hubbard model Eq.~(\ref{HubbardModel}) in the strongly interacting limit with $t \ll T \ll U$ \cite{oitmaa}. Splitting explicitly the potential term for the two families and including the chemical potentials $\mu_{A,B}$ as in Eq.~(\ref{EnergyFunction}), we first write the Hamiltonian (\ref{HubbardModel}) as
\beqa
\label{TwoVarietyNColorHamiltonian}
H(t) &=&  \frac{U}{2} \sum \limits_j \hat{n}_j (\hat{n}_j - 1) + \sum \limits_{j,\alpha \in A} (V_{A,j} - \mu_A) \hat{n}_{\alpha,j} \nonumber \\
&& + \sum \limits_{j,\alpha \in B} (V_{B,j} - \mu_B) \hat{n}_{\alpha,j} - t \sum \limits_{\braket{jk},\alpha} c^{\dagger}_{\alpha,j} c_{\alpha,k} \nonumber \\
&=& H_0 - t \, {\cal T}
\eeqa
Having denoted the hopping operator as ${\cal T} = \sum_{\braket{jk},\alpha} c^{\dagger}_{\alpha,j} c_{\alpha,k}$, the lowest non-trivial term contributing to the grand potential $\Omega$ is second order in the small expansion parameter $t$ and is given by
\begin{equation}
\label{secondOrderTerm_BeforeIntegral}
- \beta \Omega_2 = t^2 \int_{0}^{\beta} \mathrm{d} \tau_1 \, \int_{0}^{\tau_1} \mathrm{d} \tau_2 \, \langle \tilde{\mathcal{T}}(\tau_1) \tilde{\mathcal{T}}(\tau_2) \rangle_{L},
\end{equation}
where $\tilde{\mathcal{T}}(\tau) = e^{\tau H_0} \mathcal{T} e^{- \tau H_0}$, $\braket{ \hat{O} } = \text{Tr} \, [e^{-\beta H_0} \hat{O}] / \text{Tr} \, [e^{-\beta H_0}]$ is the expectation value of operator $\hat{O}$ with respect to the atomic limit Hamiltonian $H_0$ and $\braket{O}_L$ stands for the term in $\braket{O}$ proportional to the number of sites $L$, see \cite{Henderson_1992_PRB} and chapter 8 of \cite{oitmaa} for details.

In the atomic limit $H_0 = H(t=0) = \sum_{j=1}^L h_{0j}$ is a sum of Hamiltonians acting only on a single site $j$ of the system. Similarly, ${\cal T}$ connects only nearest-neighbor sites which differ by a single particle of color $\alpha$. In this case, two such nearest-neighbor sites (denoted by 1 and 2 hereafter) are spanned by eigenvectors of $H_0$ $\ket{m_{12}} = \ket{m_1} \ket{m_2}$ with the corresponding eigenenergy $E_{m_{12}} = \braket{m_{12} | H_0 | m_{12}} = \epsilon_{m_1} + \epsilon_{m_2}$, where the single-site energies $\epsilon_{m_j}$ are given by Eq.~\eqref{EnergyFunction}.
Using this and the LDA, the Eq.~(\ref{secondOrderTerm_BeforeIntegral}) can be written as $-\beta \Omega_2 = -\beta \sum_j \Omega_{2,j}$, where
\beqa
	-\beta \Omega_{2,j} &=& t^2 c_\ell z_0^{-2} \sum \limits_{m_{12},p_{12}} e^{-\beta (\epsilon_{m_1} + \epsilon_{m_2}) } |\langle p_{12} | {\cal T} | m_{12}\rangle|^2 \times \nonumber \\
&& \phantom{==} \times I(\epsilon_{m_1} + \epsilon_{m_2} - \epsilon_{p_1} - \epsilon_{p_2}),
\label{eq:sum_SecondOrder}
\eeqa
$c_\ell$ is the coordination number of the lattice, $z_0$ the single-site partition function Eq.~\eqref{TwoVarietySum} and
\beqa
\label{theIntegralFunction_app}
	I(\Delta) &=& \int_0^\beta {\rm d}\tau_1 \, \int_0^{\tau_1} {\rm d}\tau_2 \, {\rm e}^{\tau_1 \Delta} {\rm e}^{\tau_2 \Delta} \nonumber \\
	&=&
	\begin{cases}
	\frac{\beta^2}{2} & , \Delta = 0\\
	\frac{1}{\Delta^2}(e^{\beta \Delta} - \beta \Delta - 1) & , \Delta \neq 0.
	\end{cases}
\eeqa
with the result stated in Eq.~\eqref{theIntegralFunction}.

The sum in (\ref{eq:sum_SecondOrder}) can be evaluated as follows. Let us denote the number of particles of family $F$ and its complement $\bar{F}$ on site 1 and 2 as $n_{1F}, n_{2F}, n_{1\bar{F}}, n_{2\bar{F}}$ respectively. Next, we consider a hopping of a particle of the family $F$ from site 1 to site 2. The only non-vanishing contribution to the sum (\ref{eq:sum_SecondOrder}) comes from a configuration where there is exactly one particle of color $\alpha \in F$ on site 1 and zero such particles on site 2. We can choose the color $\alpha$ on site 1 from $N_F$ possibilities. The remaining $n_{1F}-1$ particles of family $F$ on site 1 can be chosen in $\binom{N_F-1}{n_{1F}-1}$ ways. Similarly, there are $\binom{N_F-1}{n_{2F}}$ possible configurations of particles of family $F$ on site 2. The number of configurations of particles belonging to the complementary family $\bar{F}$ is not constrained by the configurations of the family $F$ and is given by $\binom{N_{\bar{F}}}{n_{1\bar{F}}}$, $\binom{N_{\bar{F}}}{n_{2\bar{F}}}$ on site 1 and 2 respectively. The overall combinatorial factor is thus the product of all these factors, namely
\begin{equation}
	N_F \binom{N_F-1}{n_{1F}-1} \binom{N_F-1}{n_{2F}} \binom{N_{\bar{F}}}{n_{1\bar{F}}} \binom{N_{\bar{F}}}{n_{2\bar{F}}},
\label{fig:TwoVarietyColorMatrixElement}
\end{equation}
which appears in the Eq.~(\ref{TwoVarietySecondOrderTerm}). We also note that to convert the sum over $m_{12},p_{12}$ in the Eq.~(\ref{eq:sum_SecondOrder}) to a sum over $n_{1F},n_{2F},n_{1\bar{F}},n_{2\bar{F}}$, we have exploited the fact that the single-site energies $\epsilon_{m_j} = \epsilon_{m_j}(n_{jA},n_{jB})$, Eq.~(\ref{EnergyFunction}), are only functions of $n_{jF},n_{j\bar{F}}$.

\section{Extrema of the entropy density}
\label{app:Symmetry}

In this section we show by explicit computation in the atomic limit and in the regime of small particle density in the reservoir, $\bar{n}_{AR} + \bar{n}_{BR} < 1$, that the symmetric choice of chemical potentials $\mu_A = \mu_B$ for the two families corresponds to the extremum of the entropy density per particle
\beqa
\label{eq:s_density_app}
	\bar{s} = \bar{s}_i &=& \frac{L_R s_R + L_D s_D}{L_R(\bar{n}_{AR} + \bar{n}_{BR}) + L_D (\bar{n}_{AD} + \bar{n}_{BD})} \nonumber \\
	&=& \frac{s_R + r s_D}{{\frak n} + r {\frak n}_D} \nonumber \\
	&=:& \frac{Y}{W}.
\eeqa
investigated in Fig.~{\ref{fig:entropy_isolines}}a. Here ${\frak n} = \sum_{F=A,B} \bar{n}_{FR}$, ${\frak n}_D = \sum_{F=A,B} \bar{n}_{FD}$ and $r=L_D/L_R$ is the ratio of the dimple and the reservoir sizes. 
The functions $Y,W$ in (\ref{eq:s_density_app}) stand for the nominator and the denominator respectively and are defined for future convenience, cf. below.

In the limit of zero tunneling (atomic limit), large interactions, $\beta U \gg 1$ and $\mu_F < U$, the dominant contribution to the single-site partition function in the reservoir comes from the configurations containing at most one particle such that the Eq.~(\ref{TwoVarietySum}) can be approximated as
\beq
	z_0 \approx 1 + \sum_F N_F {\rm e}^{\beta \mu_F},
	\label{eq:z0_app}
\eeq
where we have used the fact that $V_{A,j} = V_{B,j}=0$ (we drop the site index hereafter for simplicity as we will be concerned solely with the quantities in the reservoir and the atomic limit; we also use $F=A,B$ and for a given $F$ we denote its complement as $\bar{F}$ throughout this section). The corresponding particle and entropy densities (\ref{eq:ni}),(\ref{eq:si}) read
\beqa
	\bar{n}_F &=& \frac{1}{z_0} N_F {\rm e}^{\beta \mu_F} \label{eq:barnF_app}\\
	s &=& \log(z_0) - \frac{\beta}{z_0} \sum_F \mu_F N_F {\rm e}^{\beta \mu_F}. \label{eq:s_app}
\eeqa
From (\ref{eq:barnF_app}) we find ${\rm e}^{\beta \mu_F} N_F = \bar{n}_F z_0$ which allows to express the partition function (\ref{eq:z0_app}) as
\beq
	z_0 = \frac{1}{1-{\frak n}} \label{eq:z0_simple}
\eeq
and consequently the entropy density (\ref{eq:s_app}) as 
\beq
	s = -\log(1-\frak{n}) - \sum_F \beta \mu_F {\bar n}_F.
\eeq
It is interesting to verify that combining (\ref{eq:barnF_app}) and (\ref{eq:z0_simple}) we also get
\beq
	{\frak n}^2 - {\frak n} + N {\rm e}^{\beta \mu} = 0
	\label{eq:frakn}
\eeq
which has real solutions only on the interval $0 \leq {\mathfrak n} \leq 1$ consistently with the approximate expressions for the on-site partition function (\ref{eq:z0_app}) which neglects contributions from larger particle densities (we recall that  $N=N_A + N_B$ is the total number of colors). 

Next, we assume that the entropy and particle densities in the dimple $s_D, \bar{n}_{FD}$ do not vary with the chemical potentials $\mu_F$, which is well satisfied when the dimple is in the Mott regime (we further comment on this assumption below). In what follows, we investigate the extrema of the reservoir density Eq.~(\ref{eq:s_density_app}) with respect to $\mu_F$. Denoting $\partial \equiv \partial_{\mu_F}, \bar{\partial} \equiv \partial_{\mu_{\bar F}}$ to simplify the notation,  the extremum has to satisfy $\partial \bar{s} = \bar{\partial} \bar{s} = 0$. Applying this condition to the Eq.~(\ref{eq:s_density_app}) we find 
\beq
	\partial \bar{s} = 0 \; \Leftrightarrow \; W \partial Y - Y \partial W = 0
	\label{eq:extremum}
\eeq
which yields the constraint for the values of $\mu_A, \mu_B$ extremizing $\bar{s}$. Using
\begin{subequations}
	\label{eq:relations_n}
	\begin{align}
		\partial \bar{n}_F &= \beta (1-\bar{n}_F) \bar{n}_F \\
		\bar{\partial} \bar{n}_F &= - \beta \bar{n}_F \bar{n}_{\bar F} \\
		\partial z_0 &= \beta z_0 \bar{n}_F.
	\end{align}
\end{subequations}
we have
\begin{subequations}
	\begin{align}
		\partial Y &= \beta \bar{n}_F \left[ \beta \mu_F (\bar{n}_F - 1) + \beta \mu_{\bar{F}} \bar{n}_{\bar{F}} \right] \\
		\partial W &= \beta \bar{n}_F (1-\frak{n}).
	\end{align}
\end{subequations}

To proceed, rather than investigating the properties of the constraint (\ref{eq:extremum}) for the general variables $\mu_A, \mu_B$, we ask whether it can be satisfied for $\mu_A = \mu_B = \mu$. In this case
\begin{subequations}
	\label{eq:expressions}
	\begin{align}
		{\frak n} &= \frac{N}{N_F} \bar{n}_F \label{eq:frakn_n}\\
		\bar{n}_F &= \frac{N_F}{N_{\bar F}} \bar{n}_{\bar F} \label{eq:n_n}\\
		\beta\mu & = \log\left(\frac{1}{N}\frac{\frak n}{1-{\frak n}} \right). \label{eq:beta_mu}
	\end{align}
\end{subequations}
Substituting these expressions to (\ref{eq:extremum}) we find
\beq
	\beta \bar{n}_F \left[ r(\frak{n}_D \beta \mu + s_D) + \frak{n} \beta \mu + s \right] = 0.
\eeq
The first solution is, with the help of (\ref{eq:frakn_n}), the trivial limit $\frak{n}=0$, i.e. vanishing particle density in the reservoir. The second solution can be cast in the form
\beq
	\frac{P}{Q} = r,
	\label{eq:PQ}
\eeq
where
\begin{subequations}
	\begin{align}
		P &=-(\frak{n} \beta \mu + s) = {\rm log}(1-\frak{n})  \\
		Q &= \frak{n}_D \beta \mu + s_D = \frak{n}_D \, {\rm log}\left( \frac{\eta_D}{N} \frac{\frak{n}}{1-\frak{n}} \right).
	\end{align}
\end{subequations}
Here $\log \eta_D = s_D/{\frak n}_D$ and we have used the expression (\ref{eq:beta_mu}) for $\beta \mu$. For a given dimple to reservoir size ratio $r$ the Eq.~(\ref{eq:PQ}) thus represents the condition for ${\frak n}$, and through (\ref{eq:frakn_n}) for $\bar{n}_F$ and $\bar{n}_{\bar F}$, which maximizes $\bar{s}$.
For the physically meaningful scenario ${\frak n}_D >0$ we find that for $\frak{n} \in (0,1)$, cf. the Eq.~(\ref{eq:frakn}), $P \in (-\infty,0)$ and $Q \in (-\infty,\infty)$ with the limit $\lim_{\frak{n} \to 0^+} P = 0$. This implies that the condition (\ref{eq:PQ}) can be satisfied for arbitrary $r$ for $0 < \frak{n} < 1$, proving that $\mu_A = \mu_B$ corresponds to the extremum of $\bar{s}$ in the atomic limit as claimed.

To demonstrate this, we consider the case studied in Fig.~\ref{fig:entropy_isolines}a, where $r=1/50$ and ${\frak n}_D = \bar{n}_{AD}=1$ such that $s_D = {\rm log} N_A$ and thus $\eta_D = N_A$. Solving numerically the Eq.~(\ref{eq:PQ}) and using (\ref{eq:frakn_n}),(\ref{eq:n_n}) we get for the maximum $(\bar{n}_A, \bar{n}_B) \approx (0.016, 0.063)$ in agreement with the Fig.~\ref{fig:entropy_isolines}a.

To conclude, we remark that the upper limit ${\frak n} = 1$ corresponds to the boundary delimiting the Mott regimes in the dimple the particle densities of which differ by one, cf. the inset in Fig.~{\ref{fig:entropy_isolines}}a with $\bar{n}_{BR} = 1 - \bar{n}_{AR}$ delimiting regions of $\bar{n}_{AD}=1$ and $\bar{n}_{AD}=2$ respectively.

\section{Benchmarking the second order high-temperature expansion against DMFT}

\label{app:DMFT}

The high-temperature expansion of the Hubbard model is appealing due to its relative simplicity, however its validity is limited, as the name suggests, to high temperatures $T_f \gtrsim t$ \cite{Jordens_2010_PRL}. While the use of advanced numerical methods to address low temperatures goes beyond the scope of the present work, cf. also the discussion in Sec.~\ref{sec:Results} and Sec.~\ref{sec:Conclusions}, here we compare the second order high-temperature expansion against existing DMFT data of Ref.~\cite{Bernier_2009_PRA} for a ${\rm SU}(2)$ Hubbard model with a dimple. This is a scenario which is conceptually equivalent to the present study.

Based on \cite{Bernier_2009_PRA}, we consider a Hubbard model with a three-dimensional, rotationally symmetric potential $V(r,z) = V_\text{harmonic} + V_\text{dimple} + V_\text{barrier} + V_0$, with

\begin{subequations}
\label{eq:V_dimple3D}
\begin{align}
V_\text{harmonic}(r,z) & = V_h \, (r^2 + \gamma^2 z^2) / a \\
V_\text{dimple}(r,z) & = -V_d \, \text{exp}\left (-2r^2 / w_d^2 \right ) \\
V_\text{barrier}(r,z) & = V_b \, \text{exp} \left ( -2 (r - r_b)^2 / w_b^2 \right ) 
\end{align}
\end{subequations}
and parameters $\gamma^2 = 50$, $V_h / 6t = 1.8 \times 10^{-4}$, $V_b / 6t = 6$, $V_d / 6t = 15$, $r_b = 15a$, $w_b = 5a$, $w_d = 15a$, and $a$ is the lattice spacing.
As the system is three-dimensional, the coordination number $c_\ell = 6$ for a simple cubic lattice.
The offset $V_0$ is chosen such that $V(0,0) = 0$.

\begin{figure}
  \centering
  	\includegraphics[width=0.5\textwidth]{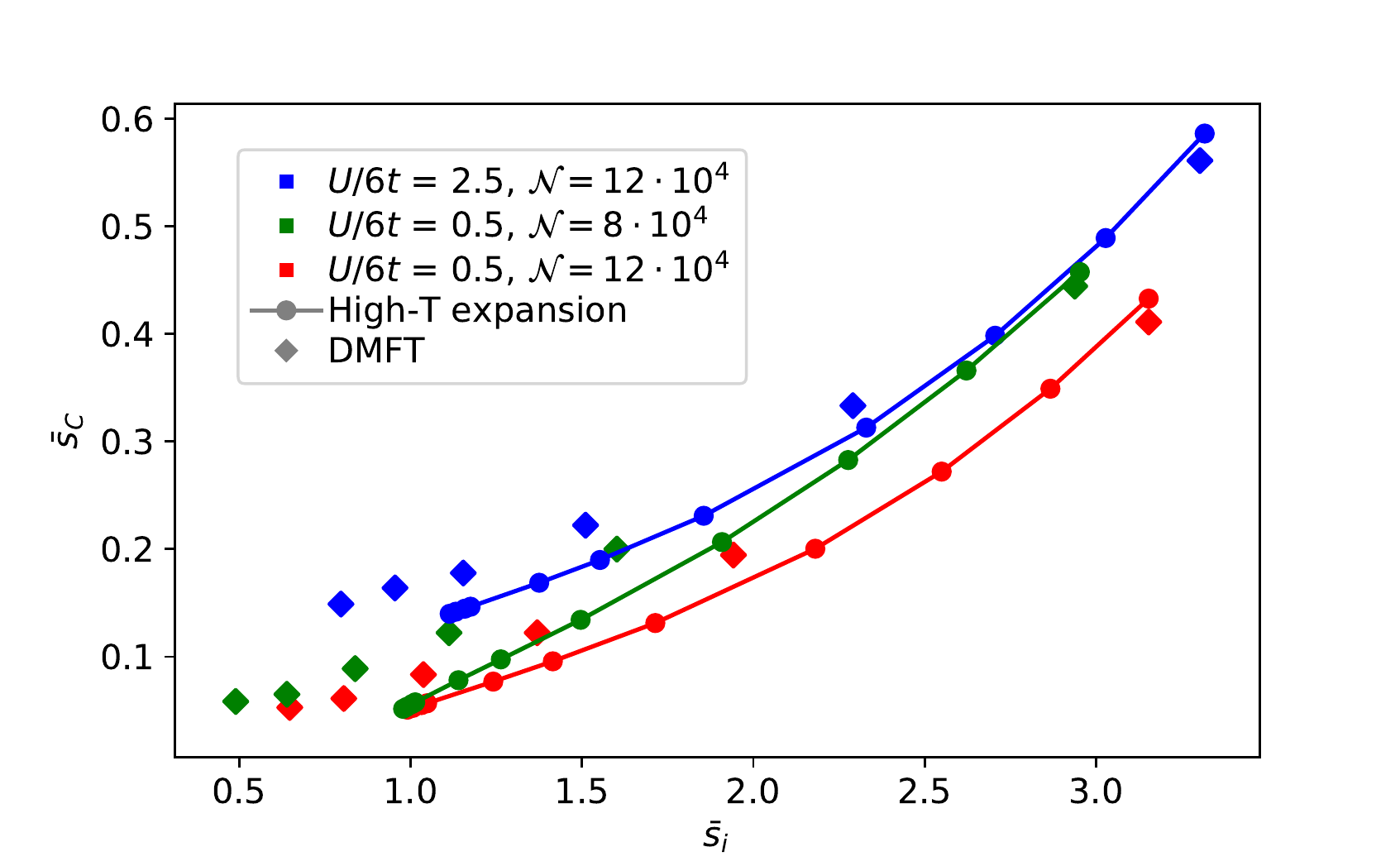}
	\caption{
	Entropy density per particle $\bar{s}_C$ in the center of the dimple (\ref{eq:V_dimple3D}) as a function of the initial entropy density per particle $\bar{s}_i$. The solid circles connected by lines are obtained using second-order high-temperature expansion of the Hubbard model. The diamonds are the DMFT data taken from Fig. 3b of Ref. \cite{Bernier_2009_PRA}. Blue, green and red data correspond to the interaction strengths $U/6t=2.5,0.5,0.5$ and total number of particles ${\cal N} = (12,8,12) \cdot 10^4$ respectively.
	}
	\label{fig:BernierBenchmark}
\end{figure}

Using the high-temperature expansion to second order within LDA, we evaluate the entropy density per particle $\bar{s}_C$ in the dimple ($r < r_b$) as a function of the initial entropy density per particle  $\bar{s}_i$. The results (solid circles and lines) for different values of the interaction strength and total number of particles $\cal N$ are shown in Fig.~\ref{fig:BernierBenchmark}, where they are compared with the DMFT results (diamonds) extracted from Fig.~3b of \cite{Bernier_2009_PRA}. For all data, we find reasonable agreement which improves with increasing $\bar{s}_i$ (increasing $T_i$). Furthermore, the data agree semi-quantitatively (within a factor of 2) in the limit of low final temperatures $T_f \approx t$ corresponding to the region with $\bar{s}_i \approx 1$ in Fig.~\ref{fig:BernierBenchmark}.

We note that a similar comparison between the DMFT and high-temperature expansions (up to the $10^{\rm th}$ order) of the Hubbard model has been performed in Ref.~\cite{Jordens_2010_PRL} which reached identical conclusion, namely that the high-temperature expansion agrees with the DMFT for temperatures down to $T_f \gtrsim t$. As in the main text we consider $T_f > 2t$, the agreement shown in Fig.~\ref{fig:BernierBenchmark} is a strong indication of the reasonable quantitative accuracy of the high-temperature expansion used in the present context of the two-family Hubbard model.

\end{document}

%% file: SUN_Cooling.bbl
%

%% file: SUN_Cooling.bbl
\begin{thebibliography}{73}%
\makeatletter
\providecommand \@ifxundefined [1]{%
 \@ifx{#1\undefined}
}%
\providecommand \@ifnum [1]{%
 \ifnum #1\expandafter \@firstoftwo
 \else \expandafter \@secondoftwo
 \fi
}%
\providecommand \@ifx [1]{%
 \ifx #1\expandafter \@firstoftwo
 \else \expandafter \@secondoftwo
 \fi
}%
\providecommand \natexlab [1]{#1}%
\providecommand \enquote  [1]{``#1''}%
\providecommand \bibnamefont  [1]{#1}%
\providecommand \bibfnamefont [1]{#1}%
\providecommand \citenamefont [1]{#1}%
\providecommand \href@noop [0]{\@secondoftwo}%
\providecommand \href [0]{\begingroup \@sanitize@url \@href}%
\providecommand \@href[1]{\@@startlink{#1}\@@href}%
\providecommand \@@href[1]{\endgroup#1\@@endlink}%
\providecommand \@sanitize@url [0]{\catcode `\\12\catcode `\$12\catcode
  `\&12\catcode `\#12\catcode `\^12\catcode `\_12\catcode `\%12\relax}%
\providecommand \@@startlink[1]{}%
\providecommand \@@endlink[0]{}%
\providecommand \url  [0]{\begingroup\@sanitize@url \@url }%
\providecommand \@url [1]{\endgroup\@href {#1}{\urlprefix }}%
\providecommand \urlprefix  [0]{URL }%
\providecommand \Eprint [0]{\href }%
\providecommand \doibase [0]{http://dx.doi.org/}%
\providecommand \selectlanguage [0]{\@gobble}%
\providecommand \bibinfo  [0]{\@secondoftwo}%
\providecommand \bibfield  [0]{\@secondoftwo}%
\providecommand \translation [1]{[#1]}%
\providecommand \BibitemOpen [0]{}%
\providecommand \bibitemStop [0]{}%
\providecommand \bibitemNoStop [0]{.\EOS\space}%
\providecommand \EOS [0]{\spacefactor3000\relax}%
\providecommand \BibitemShut  [1]{\csname bibitem#1\endcsname}%
\let\auto@bib@innerbib\@empty
\bibitem [{\citenamefont {Esslinger}(2010)}]{Esslinger_2010}%
  \BibitemOpen
  \bibfield  {author} {\bibinfo {author} {\bibfnamefont {T.}~\bibnamefont
  {Esslinger}},\ }\href@noop {} {\bibfield  {journal} {\bibinfo  {journal}
  {Annu. Rev. Condens. Matter Phys.}\ }\textbf {\bibinfo {volume} {1}},\
  \bibinfo {pages} {129} (\bibinfo {year} {2010})}\BibitemShut {NoStop}%
\bibitem [{\citenamefont {Fukuhara}\ \emph {et~al.}(2007)\citenamefont
  {Fukuhara}, \citenamefont {Takasu}, \citenamefont {Kumakura},\ and\
  \citenamefont {Takahashi}}]{Fukuhara_2007_PRL}%
  \BibitemOpen
  \bibfield  {author} {\bibinfo {author} {\bibfnamefont {T.}~\bibnamefont
  {Fukuhara}}, \bibinfo {author} {\bibfnamefont {Y.}~\bibnamefont {Takasu}},
  \bibinfo {author} {\bibfnamefont {M.}~\bibnamefont {Kumakura}}, \ and\
  \bibinfo {author} {\bibfnamefont {Y.}~\bibnamefont {Takahashi}},\ }\href
  {\doibase 10.1103/PhysRevLett.98.030401} {\bibfield  {journal} {\bibinfo
  {journal} {Phys. Rev. Lett.}\ }\textbf {\bibinfo {volume} {98}},\ \bibinfo
  {pages} {030401} (\bibinfo {year} {2007})}\BibitemShut {NoStop}%
\bibitem [{\citenamefont {Taie}\ \emph {et~al.}(2010)\citenamefont {Taie},
  \citenamefont {Takasu}, \citenamefont {Sugawa}, \citenamefont {Yamazaki},
  \citenamefont {Tsujimoto}, \citenamefont {Murakami},\ and\ \citenamefont
  {Takahashi}}]{Taie_2010_PRL}%
  \BibitemOpen
  \bibfield  {author} {\bibinfo {author} {\bibfnamefont {S.}~\bibnamefont
  {Taie}}, \bibinfo {author} {\bibfnamefont {Y.}~\bibnamefont {Takasu}},
  \bibinfo {author} {\bibfnamefont {S.}~\bibnamefont {Sugawa}}, \bibinfo
  {author} {\bibfnamefont {R.}~\bibnamefont {Yamazaki}}, \bibinfo {author}
  {\bibfnamefont {T.}~\bibnamefont {Tsujimoto}}, \bibinfo {author}
  {\bibfnamefont {R.}~\bibnamefont {Murakami}}, \ and\ \bibinfo {author}
  {\bibfnamefont {Y.}~\bibnamefont {Takahashi}},\ }\href {\doibase
  10.1103/PhysRevLett.105.190401} {\bibfield  {journal} {\bibinfo  {journal}
  {Phys. Rev. Lett.}\ }\textbf {\bibinfo {volume} {105}},\ \bibinfo {pages}
  {190401} (\bibinfo {year} {2010})}\BibitemShut {NoStop}%
\bibitem [{\citenamefont {Sugawa}\ \emph {et~al.}(2013)\citenamefont {Sugawa},
  \citenamefont {Takasu}, \citenamefont {Enomoto},\ and\ \citenamefont
  {Takahashi}}]{sugawa2013ultracold}%
  \BibitemOpen
  \bibfield  {author} {\bibinfo {author} {\bibfnamefont {S.}~\bibnamefont
  {Sugawa}}, \bibinfo {author} {\bibfnamefont {Y.}~\bibnamefont {Takasu}},
  \bibinfo {author} {\bibfnamefont {K.}~\bibnamefont {Enomoto}}, \ and\
  \bibinfo {author} {\bibfnamefont {Y.}~\bibnamefont {Takahashi}},\ }in\
  \href@noop {} {\emph {\bibinfo {booktitle} {Annual Review of Cold Atoms and
  Molecules: Volume 1}}}\ (\bibinfo  {publisher} {World Scientific},\ \bibinfo
  {year} {2013})\ pp.\ \bibinfo {pages} {3--51}\BibitemShut {NoStop}%
\bibitem [{\citenamefont {DeSalvo}\ \emph {et~al.}(2010)\citenamefont
  {DeSalvo}, \citenamefont {Yan}, \citenamefont {Mickelson}, \citenamefont
  {Escobar},\ and\ \citenamefont {Killian}}]{DeSalvo_2010_PRL}%
  \BibitemOpen
  \bibfield  {author} {\bibinfo {author} {\bibfnamefont {B.}~\bibnamefont
  {DeSalvo}}, \bibinfo {author} {\bibfnamefont {M.}~\bibnamefont {Yan}},
  \bibinfo {author} {\bibfnamefont {P.}~\bibnamefont {Mickelson}}, \bibinfo
  {author} {\bibfnamefont {Y.}~\bibnamefont {Escobar}}, \ and\ \bibinfo
  {author} {\bibfnamefont {T.}~\bibnamefont {Killian}},\ }\href {\doibase
  10.1103/PhysRevLett.105.030402} {\bibfield  {journal} {\bibinfo  {journal}
  {Physical Review Letters}\ }\textbf {\bibinfo {volume} {105}},\ \bibinfo
  {pages} {030402} (\bibinfo {year} {2010})}\BibitemShut {NoStop}%
\bibitem [{\citenamefont {Stellmer}\ \emph {et~al.}(2013)\citenamefont
  {Stellmer}, \citenamefont {Grimm},\ and\ \citenamefont
  {Schreck}}]{Stellmer_2013_PRA}%
  \BibitemOpen
  \bibfield  {author} {\bibinfo {author} {\bibfnamefont {S.}~\bibnamefont
  {Stellmer}}, \bibinfo {author} {\bibfnamefont {R.}~\bibnamefont {Grimm}}, \
  and\ \bibinfo {author} {\bibfnamefont {F.}~\bibnamefont {Schreck}},\ }\href
  {\doibase 10.1103/PhysRevA.87.013611} {\bibfield  {journal} {\bibinfo
  {journal} {Phys. Rev. A}\ }\textbf {\bibinfo {volume} {87}},\ \bibinfo
  {pages} {013611} (\bibinfo {year} {2013})}\BibitemShut {NoStop}%
\bibitem [{\citenamefont {Stellmer}\ \emph {et~al.}(2014)\citenamefont
  {Stellmer}, \citenamefont {Schreck},\ and\ \citenamefont
  {Killian}}]{stellmer2014degenerate}%
  \BibitemOpen
  \bibfield  {author} {\bibinfo {author} {\bibfnamefont {S.}~\bibnamefont
  {Stellmer}}, \bibinfo {author} {\bibfnamefont {F.}~\bibnamefont {Schreck}}, \
  and\ \bibinfo {author} {\bibfnamefont {T.~C.}\ \bibnamefont {Killian}},\ }in\
  \href@noop {} {\emph {\bibinfo {booktitle} {Annual Review of Cold Atoms and
  Molecules}}}\ (\bibinfo  {publisher} {World Scientific},\ \bibinfo {year}
  {2014})\ pp.\ \bibinfo {pages} {1--80}\BibitemShut {NoStop}%
\bibitem [{\citenamefont {Cazalilla}\ and\ \citenamefont
  {Rey}(2014)}]{cazalilla2014ultracold}%
  \BibitemOpen
  \bibfield  {author} {\bibinfo {author} {\bibfnamefont {M.~A.}\ \bibnamefont
  {Cazalilla}}\ and\ \bibinfo {author} {\bibfnamefont {A.~M.}\ \bibnamefont
  {Rey}},\ }\href@noop {} {\bibfield  {journal} {\bibinfo  {journal} {Reports
  on Progress in Physics}\ }\textbf {\bibinfo {volume} {77}},\ \bibinfo {pages}
  {124401} (\bibinfo {year} {2014})}\BibitemShut {NoStop}%
\bibitem [{\citenamefont {Wang}\ \emph {et~al.}(2014)\citenamefont {Wang},
  \citenamefont {Li}, \citenamefont {Cai}, \citenamefont {Zhou}, \citenamefont
  {Wang},\ and\ \citenamefont {Wu}}]{wang2014competing}%
  \BibitemOpen
  \bibfield  {author} {\bibinfo {author} {\bibfnamefont {D.}~\bibnamefont
  {Wang}}, \bibinfo {author} {\bibfnamefont {Y.}~\bibnamefont {Li}}, \bibinfo
  {author} {\bibfnamefont {Z.}~\bibnamefont {Cai}}, \bibinfo {author}
  {\bibfnamefont {Z.}~\bibnamefont {Zhou}}, \bibinfo {author} {\bibfnamefont
  {Y.}~\bibnamefont {Wang}}, \ and\ \bibinfo {author} {\bibfnamefont
  {C.}~\bibnamefont {Wu}},\ }\href@noop {} {\bibfield  {journal} {\bibinfo
  {journal} {Physical Review Letters}\ }\textbf {\bibinfo {volume} {112}},\
  \bibinfo {pages} {156403} (\bibinfo {year} {2014})}\BibitemShut {NoStop}%
\bibitem [{\citenamefont {Barbarino}\ \emph {et~al.}(2015)\citenamefont
  {Barbarino}, \citenamefont {Taddia}, \citenamefont {Rossini}, \citenamefont
  {Mazza},\ and\ \citenamefont {Fazio}}]{barbarino2015magnetic}%
  \BibitemOpen
  \bibfield  {author} {\bibinfo {author} {\bibfnamefont {S.}~\bibnamefont
  {Barbarino}}, \bibinfo {author} {\bibfnamefont {L.}~\bibnamefont {Taddia}},
  \bibinfo {author} {\bibfnamefont {D.}~\bibnamefont {Rossini}}, \bibinfo
  {author} {\bibfnamefont {L.}~\bibnamefont {Mazza}}, \ and\ \bibinfo {author}
  {\bibfnamefont {R.}~\bibnamefont {Fazio}},\ }\href@noop {} {\bibfield
  {journal} {\bibinfo  {journal} {Nature communications}\ }\textbf {\bibinfo
  {volume} {6}},\ \bibinfo {pages} {1} (\bibinfo {year} {2015})}\BibitemShut
  {NoStop}%
\bibitem [{\citenamefont {Chen}\ \emph {et~al.}(2016)\citenamefont {Chen},
  \citenamefont {Hazzard}, \citenamefont {Rey},\ and\ \citenamefont
  {Hermele}}]{chen2016synthetic}%
  \BibitemOpen
  \bibfield  {author} {\bibinfo {author} {\bibfnamefont {G.}~\bibnamefont
  {Chen}}, \bibinfo {author} {\bibfnamefont {K.~R.}\ \bibnamefont {Hazzard}},
  \bibinfo {author} {\bibfnamefont {A.~M.}\ \bibnamefont {Rey}}, \ and\
  \bibinfo {author} {\bibfnamefont {M.}~\bibnamefont {Hermele}},\ }\href@noop
  {} {\bibfield  {journal} {\bibinfo  {journal} {Physical Review A}\ }\textbf
  {\bibinfo {volume} {93}},\ \bibinfo {pages} {061601} (\bibinfo {year}
  {2016})}\BibitemShut {NoStop}%
\bibitem [{\citenamefont {Capponi}\ \emph {et~al.}(2016)\citenamefont
  {Capponi}, \citenamefont {Lecheminant},\ and\ \citenamefont
  {Totsuka}}]{capponi2016phases}%
  \BibitemOpen
  \bibfield  {author} {\bibinfo {author} {\bibfnamefont {S.}~\bibnamefont
  {Capponi}}, \bibinfo {author} {\bibfnamefont {P.}~\bibnamefont
  {Lecheminant}}, \ and\ \bibinfo {author} {\bibfnamefont {K.}~\bibnamefont
  {Totsuka}},\ }\href@noop {} {\bibfield  {journal} {\bibinfo  {journal}
  {Annals of Physics}\ }\textbf {\bibinfo {volume} {367}},\ \bibinfo {pages}
  {50} (\bibinfo {year} {2016})}\BibitemShut {NoStop}%
\bibitem [{\citenamefont {Jen}\ and\ \citenamefont {Yip}(2018)}]{jen2018spin}%
  \BibitemOpen
  \bibfield  {author} {\bibinfo {author} {\bibfnamefont {H.-H.}\ \bibnamefont
  {Jen}}\ and\ \bibinfo {author} {\bibfnamefont {S.-K.}\ \bibnamefont {Yip}},\
  }\href@noop {} {\bibfield  {journal} {\bibinfo  {journal} {Physical Review
  A}\ }\textbf {\bibinfo {volume} {98}},\ \bibinfo {pages} {013623} (\bibinfo
  {year} {2018})}\BibitemShut {NoStop}%
\bibitem [{\citenamefont {Chung}\ and\ \citenamefont
  {Corboz}(2019)}]{Chung_2019_PRB}%
  \BibitemOpen
  \bibfield  {author} {\bibinfo {author} {\bibfnamefont {S.~S.}\ \bibnamefont
  {Chung}}\ and\ \bibinfo {author} {\bibfnamefont {P.}~\bibnamefont {Corboz}},\
  }\href {\doibase 10.1103/PhysRevB.100.035134} {\bibfield  {journal} {\bibinfo
   {journal} {Phys. Rev. B}\ }\textbf {\bibinfo {volume} {100}},\ \bibinfo
  {pages} {035134} (\bibinfo {year} {2019})}\BibitemShut {NoStop}%
\bibitem [{\citenamefont {Bl{\"u}mer}\ and\ \citenamefont
  {Gorelik}(2013)}]{blumer2013mott}%
  \BibitemOpen
  \bibfield  {author} {\bibinfo {author} {\bibfnamefont {N.}~\bibnamefont
  {Bl{\"u}mer}}\ and\ \bibinfo {author} {\bibfnamefont {E.}~\bibnamefont
  {Gorelik}},\ }\href@noop {} {\bibfield  {journal} {\bibinfo  {journal}
  {Physical Review B}\ }\textbf {\bibinfo {volume} {87}},\ \bibinfo {pages}
  {085115} (\bibinfo {year} {2013})}\BibitemShut {NoStop}%
\bibitem [{\citenamefont {Xu}\ \emph {et~al.}(2018)\citenamefont {Xu},
  \citenamefont {Barreiro}, \citenamefont {Wang},\ and\ \citenamefont
  {Wu}}]{xu2018interaction}%
  \BibitemOpen
  \bibfield  {author} {\bibinfo {author} {\bibfnamefont {S.}~\bibnamefont
  {Xu}}, \bibinfo {author} {\bibfnamefont {J.~T.}\ \bibnamefont {Barreiro}},
  \bibinfo {author} {\bibfnamefont {Y.}~\bibnamefont {Wang}}, \ and\ \bibinfo
  {author} {\bibfnamefont {C.}~\bibnamefont {Wu}},\ }\href@noop {} {\bibfield
  {journal} {\bibinfo  {journal} {Physical review letters}\ }\textbf {\bibinfo
  {volume} {121}},\ \bibinfo {pages} {167205} (\bibinfo {year}
  {2018})}\BibitemShut {NoStop}%
\bibitem [{\citenamefont {Zhou}\ \emph {et~al.}(2017)\citenamefont {Zhou},
  \citenamefont {Wang}, \citenamefont {Wu},\ and\ \citenamefont
  {Wang}}]{zhou2017finite}%
  \BibitemOpen
  \bibfield  {author} {\bibinfo {author} {\bibfnamefont {Z.}~\bibnamefont
  {Zhou}}, \bibinfo {author} {\bibfnamefont {D.}~\bibnamefont {Wang}}, \bibinfo
  {author} {\bibfnamefont {C.}~\bibnamefont {Wu}}, \ and\ \bibinfo {author}
  {\bibfnamefont {Y.}~\bibnamefont {Wang}},\ }\href@noop {} {\bibfield
  {journal} {\bibinfo  {journal} {Physical Review B}\ }\textbf {\bibinfo
  {volume} {95}},\ \bibinfo {pages} {085128} (\bibinfo {year}
  {2017})}\BibitemShut {NoStop}%
\bibitem [{\citenamefont {Lang}\ \emph {et~al.}(2013)\citenamefont {Lang},
  \citenamefont {Meng}, \citenamefont {Muramatsu}, \citenamefont {Wessel},\
  and\ \citenamefont {Assaad}}]{lang2013dimerized}%
  \BibitemOpen
  \bibfield  {author} {\bibinfo {author} {\bibfnamefont {T.~C.}\ \bibnamefont
  {Lang}}, \bibinfo {author} {\bibfnamefont {Z.~Y.}\ \bibnamefont {Meng}},
  \bibinfo {author} {\bibfnamefont {A.}~\bibnamefont {Muramatsu}}, \bibinfo
  {author} {\bibfnamefont {S.}~\bibnamefont {Wessel}}, \ and\ \bibinfo {author}
  {\bibfnamefont {F.~F.}\ \bibnamefont {Assaad}},\ }\href@noop {} {\bibfield
  {journal} {\bibinfo  {journal} {Physical Review Letters}\ }\textbf {\bibinfo
  {volume} {111}},\ \bibinfo {pages} {066401} (\bibinfo {year}
  {2013})}\BibitemShut {NoStop}%
\bibitem [{\citenamefont {Wolf}\ \emph {et~al.}(2018)\citenamefont {Wolf},
  \citenamefont {Schmidt},\ and\ \citenamefont {Rachel}}]{Wolf_2018_PRB}%
  \BibitemOpen
  \bibfield  {author} {\bibinfo {author} {\bibfnamefont {S.}~\bibnamefont
  {Wolf}}, \bibinfo {author} {\bibfnamefont {T.~L.}\ \bibnamefont {Schmidt}}, \
  and\ \bibinfo {author} {\bibfnamefont {S.}~\bibnamefont {Rachel}},\ }\href
  {\doibase 10.1103/PhysRevB.98.174515} {\bibfield  {journal} {\bibinfo
  {journal} {Phys. Rev. B}\ }\textbf {\bibinfo {volume} {98}},\ \bibinfo
  {pages} {174515} (\bibinfo {year} {2018})}\BibitemShut {NoStop}%
\bibitem [{\citenamefont {Choudhury}\ \emph {et~al.}(2020)\citenamefont
  {Choudhury}, \citenamefont {Islam}, \citenamefont {Hou}, \citenamefont
  {Aman}, \citenamefont {Killian},\ and\ \citenamefont
  {Hazzard}}]{choudhury2020collective}%
  \BibitemOpen
  \bibfield  {author} {\bibinfo {author} {\bibfnamefont {S.}~\bibnamefont
  {Choudhury}}, \bibinfo {author} {\bibfnamefont {K.~R.}\ \bibnamefont
  {Islam}}, \bibinfo {author} {\bibfnamefont {Y.}~\bibnamefont {Hou}}, \bibinfo
  {author} {\bibfnamefont {J.~A.}\ \bibnamefont {Aman}}, \bibinfo {author}
  {\bibfnamefont {T.~C.}\ \bibnamefont {Killian}}, \ and\ \bibinfo {author}
  {\bibfnamefont {K.~R.}\ \bibnamefont {Hazzard}},\ }\href@noop {} {\bibfield
  {journal} {\bibinfo  {journal} {Physical Review A}\ }\textbf {\bibinfo
  {volume} {101}},\ \bibinfo {pages} {053612} (\bibinfo {year}
  {2020})}\BibitemShut {NoStop}%
\bibitem [{\citenamefont {Pagano}\ \emph {et~al.}(2014)\citenamefont {Pagano},
  \citenamefont {Mancini}, \citenamefont {Cappellini}, \citenamefont
  {Lombardi}, \citenamefont {Sch{\"a}fer}, \citenamefont {Hu}, \citenamefont
  {Liu}, \citenamefont {Catani}, \citenamefont {Sias}, \citenamefont {Inguscio}
  \emph {et~al.}}]{pagano2014one}%
  \BibitemOpen
  \bibfield  {author} {\bibinfo {author} {\bibfnamefont {G.}~\bibnamefont
  {Pagano}}, \bibinfo {author} {\bibfnamefont {M.}~\bibnamefont {Mancini}},
  \bibinfo {author} {\bibfnamefont {G.}~\bibnamefont {Cappellini}}, \bibinfo
  {author} {\bibfnamefont {P.}~\bibnamefont {Lombardi}}, \bibinfo {author}
  {\bibfnamefont {F.}~\bibnamefont {Sch{\"a}fer}}, \bibinfo {author}
  {\bibfnamefont {H.}~\bibnamefont {Hu}}, \bibinfo {author} {\bibfnamefont
  {X.-J.}\ \bibnamefont {Liu}}, \bibinfo {author} {\bibfnamefont
  {J.}~\bibnamefont {Catani}}, \bibinfo {author} {\bibfnamefont
  {C.}~\bibnamefont {Sias}}, \bibinfo {author} {\bibfnamefont {M.}~\bibnamefont
  {Inguscio}},  \emph {et~al.},\ }\href@noop {} {\bibfield  {journal} {\bibinfo
   {journal} {Nature Physics}\ }\textbf {\bibinfo {volume} {10}},\ \bibinfo
  {pages} {198} (\bibinfo {year} {2014})}\BibitemShut {NoStop}%
\bibitem [{\citenamefont {Hofrichter}\ \emph {et~al.}(2016)\citenamefont
  {Hofrichter}, \citenamefont {Riegger}, \citenamefont {Scazza}, \citenamefont
  {H\"ofer}, \citenamefont {Fernandes}, \citenamefont {Bloch},\ and\
  \citenamefont {F\"olling}}]{PhysRevX.6.021030}%
  \BibitemOpen
  \bibfield  {author} {\bibinfo {author} {\bibfnamefont {C.}~\bibnamefont
  {Hofrichter}}, \bibinfo {author} {\bibfnamefont {L.}~\bibnamefont {Riegger}},
  \bibinfo {author} {\bibfnamefont {F.}~\bibnamefont {Scazza}}, \bibinfo
  {author} {\bibfnamefont {M.}~\bibnamefont {H\"ofer}}, \bibinfo {author}
  {\bibfnamefont {D.~R.}\ \bibnamefont {Fernandes}}, \bibinfo {author}
  {\bibfnamefont {I.}~\bibnamefont {Bloch}}, \ and\ \bibinfo {author}
  {\bibfnamefont {S.}~\bibnamefont {F\"olling}},\ }\href {\doibase
  10.1103/PhysRevX.6.021030} {\bibfield  {journal} {\bibinfo  {journal} {Phys.
  Rev. X}\ }\textbf {\bibinfo {volume} {6}},\ \bibinfo {pages} {021030}
  (\bibinfo {year} {2016})}\BibitemShut {NoStop}%
\bibitem [{\citenamefont {Ozawa}\ \emph {et~al.}(2018)\citenamefont {Ozawa},
  \citenamefont {Taie}, \citenamefont {Takasu},\ and\ \citenamefont
  {Takahashi}}]{Ozawa_2018_PRL}%
  \BibitemOpen
  \bibfield  {author} {\bibinfo {author} {\bibfnamefont {H.}~\bibnamefont
  {Ozawa}}, \bibinfo {author} {\bibfnamefont {S.}~\bibnamefont {Taie}},
  \bibinfo {author} {\bibfnamefont {Y.}~\bibnamefont {Takasu}}, \ and\ \bibinfo
  {author} {\bibfnamefont {Y.}~\bibnamefont {Takahashi}},\ }\href {\doibase
  10.1103/PhysRevLett.121.225303} {\bibfield  {journal} {\bibinfo  {journal}
  {Phys. Rev. Lett.}\ }\textbf {\bibinfo {volume} {121}},\ \bibinfo {pages}
  {225303} (\bibinfo {year} {2018})}\BibitemShut {NoStop}%
\bibitem [{\citenamefont {Taie}\ \emph {et~al.}(2020)\citenamefont {Taie},
  \citenamefont {Ibarra-Garc{\'\i}a-Padilla}, \citenamefont {Nishizawa},
  \citenamefont {Takasu}, \citenamefont {Kuno}, \citenamefont {Wei},
  \citenamefont {Scalettar}, \citenamefont {Hazzard},\ and\ \citenamefont
  {Takahashi}}]{taie2020observation}%
  \BibitemOpen
  \bibfield  {author} {\bibinfo {author} {\bibfnamefont {S.}~\bibnamefont
  {Taie}}, \bibinfo {author} {\bibfnamefont {E.}~\bibnamefont
  {Ibarra-Garc{\'\i}a-Padilla}}, \bibinfo {author} {\bibfnamefont
  {N.}~\bibnamefont {Nishizawa}}, \bibinfo {author} {\bibfnamefont
  {Y.}~\bibnamefont {Takasu}}, \bibinfo {author} {\bibfnamefont
  {Y.}~\bibnamefont {Kuno}}, \bibinfo {author} {\bibfnamefont {H.-T.}\
  \bibnamefont {Wei}}, \bibinfo {author} {\bibfnamefont {R.~T.}\ \bibnamefont
  {Scalettar}}, \bibinfo {author} {\bibfnamefont {K.~R.}\ \bibnamefont
  {Hazzard}}, \ and\ \bibinfo {author} {\bibfnamefont {Y.}~\bibnamefont
  {Takahashi}},\ }\href@noop {} {\bibfield  {journal} {\bibinfo  {journal}
  {arXiv preprint arXiv:2010.07730}\ } (\bibinfo {year} {2020})}\BibitemShut
  {NoStop}%
\bibitem [{\citenamefont {Sonderhouse}\ \emph {et~al.}(2020)\citenamefont
  {Sonderhouse}, \citenamefont {Sanner}, \citenamefont {Hutson}, \citenamefont
  {Goban}, \citenamefont {Bilitewski}, \citenamefont {Yan}, \citenamefont
  {Milner}, \citenamefont {Rey},\ and\ \citenamefont {Ye}}]{Sonderhouse2020}%
  \BibitemOpen
  \bibfield  {author} {\bibinfo {author} {\bibfnamefont {L.}~\bibnamefont
  {Sonderhouse}}, \bibinfo {author} {\bibfnamefont {C.}~\bibnamefont {Sanner}},
  \bibinfo {author} {\bibfnamefont {R.~B.}\ \bibnamefont {Hutson}}, \bibinfo
  {author} {\bibfnamefont {A.}~\bibnamefont {Goban}}, \bibinfo {author}
  {\bibfnamefont {T.}~\bibnamefont {Bilitewski}}, \bibinfo {author}
  {\bibfnamefont {L.}~\bibnamefont {Yan}}, \bibinfo {author} {\bibfnamefont
  {W.~R.}\ \bibnamefont {Milner}}, \bibinfo {author} {\bibfnamefont {A.~M.}\
  \bibnamefont {Rey}}, \ and\ \bibinfo {author} {\bibfnamefont
  {J.}~\bibnamefont {Ye}},\ }\href {\doibase 10.1038/s41567-020-0986-6}
  {\bibfield  {journal} {\bibinfo  {journal} {Nature Physics}\ } (\bibinfo
  {year} {2020}),\ 10.1038/s41567-020-0986-6}\BibitemShut {NoStop}%
\bibitem [{\citenamefont {Gorshkov}\ \emph {et~al.}(2010)\citenamefont
  {Gorshkov}, \citenamefont {Hermele}, \citenamefont {Gurarie}, \citenamefont
  {Xu}, \citenamefont {Julienne}, \citenamefont {Ye}, \citenamefont {Zoller},
  \citenamefont {Demler}, \citenamefont {Lukin},\ and\ \citenamefont
  {Rey}}]{Gorshkov_2010_NatPhys}%
  \BibitemOpen
  \bibfield  {author} {\bibinfo {author} {\bibfnamefont {A.~V.}\ \bibnamefont
  {Gorshkov}}, \bibinfo {author} {\bibfnamefont {M.}~\bibnamefont {Hermele}},
  \bibinfo {author} {\bibfnamefont {V.}~\bibnamefont {Gurarie}}, \bibinfo
  {author} {\bibfnamefont {C.}~\bibnamefont {Xu}}, \bibinfo {author}
  {\bibfnamefont {P.~S.}\ \bibnamefont {Julienne}}, \bibinfo {author}
  {\bibfnamefont {J.}~\bibnamefont {Ye}}, \bibinfo {author} {\bibfnamefont
  {P.}~\bibnamefont {Zoller}}, \bibinfo {author} {\bibfnamefont
  {E.}~\bibnamefont {Demler}}, \bibinfo {author} {\bibfnamefont {M.~D.}\
  \bibnamefont {Lukin}}, \ and\ \bibinfo {author} {\bibfnamefont
  {A.}~\bibnamefont {Rey}},\ }\href@noop {} {\bibfield  {journal} {\bibinfo
  {journal} {Nature physics}\ }\textbf {\bibinfo {volume} {6}},\ \bibinfo
  {pages} {289} (\bibinfo {year} {2010})}\BibitemShut {NoStop}%
\bibitem [{\citenamefont {Manmana}\ \emph {et~al.}(2011)\citenamefont
  {Manmana}, \citenamefont {Hazzard}, \citenamefont {Chen}, \citenamefont
  {Feiguin},\ and\ \citenamefont {Rey}}]{Manmana_2011_PRA}%
  \BibitemOpen
  \bibfield  {author} {\bibinfo {author} {\bibfnamefont {S.~R.}\ \bibnamefont
  {Manmana}}, \bibinfo {author} {\bibfnamefont {K.~R.~A.}\ \bibnamefont
  {Hazzard}}, \bibinfo {author} {\bibfnamefont {G.}~\bibnamefont {Chen}},
  \bibinfo {author} {\bibfnamefont {A.~E.}\ \bibnamefont {Feiguin}}, \ and\
  \bibinfo {author} {\bibfnamefont {A.~M.}\ \bibnamefont {Rey}},\ }\href
  {\doibase 10.1103/PhysRevA.84.043601} {\bibfield  {journal} {\bibinfo
  {journal} {Phys. Rev. A}\ }\textbf {\bibinfo {volume} {84}},\ \bibinfo
  {pages} {043601} (\bibinfo {year} {2011})}\BibitemShut {NoStop}%
\bibitem [{\citenamefont {Nataf}\ and\ \citenamefont
  {Mila}(2014)}]{nataf2014exact}%
  \BibitemOpen
  \bibfield  {author} {\bibinfo {author} {\bibfnamefont {P.}~\bibnamefont
  {Nataf}}\ and\ \bibinfo {author} {\bibfnamefont {F.}~\bibnamefont {Mila}},\
  }\href@noop {} {\bibfield  {journal} {\bibinfo  {journal} {Physical review
  letters}\ }\textbf {\bibinfo {volume} {113}},\ \bibinfo {pages} {127204}
  (\bibinfo {year} {2014})}\BibitemShut {NoStop}%
\bibitem [{\citenamefont {Nataf}\ and\ \citenamefont
  {Mila}(2016)}]{nataf2016exact}%
  \BibitemOpen
  \bibfield  {author} {\bibinfo {author} {\bibfnamefont {P.}~\bibnamefont
  {Nataf}}\ and\ \bibinfo {author} {\bibfnamefont {F.}~\bibnamefont {Mila}},\
  }\href@noop {} {\bibfield  {journal} {\bibinfo  {journal} {Physical Review
  B}\ }\textbf {\bibinfo {volume} {93}},\ \bibinfo {pages} {155134} (\bibinfo
  {year} {2016})}\BibitemShut {NoStop}%
\bibitem [{\citenamefont {Kim}\ \emph {et~al.}(2017)\citenamefont {Kim},
  \citenamefont {Penc}, \citenamefont {Nataf},\ and\ \citenamefont
  {Mila}}]{kim2017linear}%
  \BibitemOpen
  \bibfield  {author} {\bibinfo {author} {\bibfnamefont {F.~H.}\ \bibnamefont
  {Kim}}, \bibinfo {author} {\bibfnamefont {K.}~\bibnamefont {Penc}}, \bibinfo
  {author} {\bibfnamefont {P.}~\bibnamefont {Nataf}}, \ and\ \bibinfo {author}
  {\bibfnamefont {F.}~\bibnamefont {Mila}},\ }\href@noop {} {\bibfield
  {journal} {\bibinfo  {journal} {Physical Review B}\ }\textbf {\bibinfo
  {volume} {96}},\ \bibinfo {pages} {205142} (\bibinfo {year}
  {2017})}\BibitemShut {NoStop}%
\bibitem [{\citenamefont {Nataf}\ and\ \citenamefont
  {Mila}(2018)}]{nataf2018density}%
  \BibitemOpen
  \bibfield  {author} {\bibinfo {author} {\bibfnamefont {P.}~\bibnamefont
  {Nataf}}\ and\ \bibinfo {author} {\bibfnamefont {F.}~\bibnamefont {Mila}},\
  }\href@noop {} {\bibfield  {journal} {\bibinfo  {journal} {Physical Review
  B}\ }\textbf {\bibinfo {volume} {97}},\ \bibinfo {pages} {134420} (\bibinfo
  {year} {2018})}\BibitemShut {NoStop}%
\bibitem [{\citenamefont {Dufour}\ \emph {et~al.}(2015)\citenamefont {Dufour},
  \citenamefont {Nataf},\ and\ \citenamefont {Mila}}]{dufour2015variational}%
  \BibitemOpen
  \bibfield  {author} {\bibinfo {author} {\bibfnamefont {J.}~\bibnamefont
  {Dufour}}, \bibinfo {author} {\bibfnamefont {P.}~\bibnamefont {Nataf}}, \
  and\ \bibinfo {author} {\bibfnamefont {F.}~\bibnamefont {Mila}},\ }\href@noop
  {} {\bibfield  {journal} {\bibinfo  {journal} {Physical Review B}\ }\textbf
  {\bibinfo {volume} {91}},\ \bibinfo {pages} {174427} (\bibinfo {year}
  {2015})}\BibitemShut {NoStop}%
\bibitem [{\citenamefont {Romen}\ and\ \citenamefont
  {L{\"a}uchli}(2020)}]{romen2020structure}%
  \BibitemOpen
  \bibfield  {author} {\bibinfo {author} {\bibfnamefont {C.}~\bibnamefont
  {Romen}}\ and\ \bibinfo {author} {\bibfnamefont {A.~M.}\ \bibnamefont
  {L{\"a}uchli}},\ }\href@noop {} {\bibfield  {journal} {\bibinfo  {journal}
  {Physical Review Research}\ }\textbf {\bibinfo {volume} {2}},\ \bibinfo
  {pages} {043009} (\bibinfo {year} {2020})}\BibitemShut {NoStop}%
\bibitem [{\citenamefont {Chen}\ \emph {et~al.}(2015)\citenamefont {Chen},
  \citenamefont {Xue}, \citenamefont {McCulloch}, \citenamefont {Chung},
  \citenamefont {Huang},\ and\ \citenamefont {Yip}}]{chen2015quantum}%
  \BibitemOpen
  \bibfield  {author} {\bibinfo {author} {\bibfnamefont {P.}~\bibnamefont
  {Chen}}, \bibinfo {author} {\bibfnamefont {Z.-L.}\ \bibnamefont {Xue}},
  \bibinfo {author} {\bibfnamefont {I.}~\bibnamefont {McCulloch}}, \bibinfo
  {author} {\bibfnamefont {M.-C.}\ \bibnamefont {Chung}}, \bibinfo {author}
  {\bibfnamefont {C.-C.}\ \bibnamefont {Huang}}, \ and\ \bibinfo {author}
  {\bibfnamefont {S.-K.}\ \bibnamefont {Yip}},\ }\href@noop {} {\bibfield
  {journal} {\bibinfo  {journal} {Physical review letters}\ }\textbf {\bibinfo
  {volume} {114}},\ \bibinfo {pages} {145301} (\bibinfo {year}
  {2015})}\BibitemShut {NoStop}%
\bibitem [{\citenamefont {D'Emidio}\ \emph {et~al.}(2015)\citenamefont
  {D'Emidio}, \citenamefont {Block},\ and\ \citenamefont {Kaul}}]{d2015renyi}%
  \BibitemOpen
  \bibfield  {author} {\bibinfo {author} {\bibfnamefont {J.}~\bibnamefont
  {D'Emidio}}, \bibinfo {author} {\bibfnamefont {M.~S.}\ \bibnamefont {Block}},
  \ and\ \bibinfo {author} {\bibfnamefont {R.~K.}\ \bibnamefont {Kaul}},\
  }\href@noop {} {\bibfield  {journal} {\bibinfo  {journal} {Physical Review
  B}\ }\textbf {\bibinfo {volume} {92}},\ \bibinfo {pages} {054411} (\bibinfo
  {year} {2015})}\BibitemShut {NoStop}%
\bibitem [{\citenamefont {Song}\ \emph {et~al.}(2013)\citenamefont {Song},
  \citenamefont {Hermele} \emph {et~al.}}]{song2013mott}%
  \BibitemOpen
  \bibfield  {author} {\bibinfo {author} {\bibfnamefont {H.}~\bibnamefont
  {Song}}, \bibinfo {author} {\bibfnamefont {M.}~\bibnamefont {Hermele}},
  \emph {et~al.},\ }\href@noop {} {\bibfield  {journal} {\bibinfo  {journal}
  {Physical Review B}\ }\textbf {\bibinfo {volume} {87}},\ \bibinfo {pages}
  {144423} (\bibinfo {year} {2013})}\BibitemShut {NoStop}%
\bibitem [{\citenamefont {Corboz}\ \emph {et~al.}(2012)\citenamefont {Corboz},
  \citenamefont {Penc}, \citenamefont {Mila},\ and\ \citenamefont
  {L\"auchli}}]{Corboz_2012_PRB}%
  \BibitemOpen
  \bibfield  {author} {\bibinfo {author} {\bibfnamefont {P.}~\bibnamefont
  {Corboz}}, \bibinfo {author} {\bibfnamefont {K.}~\bibnamefont {Penc}},
  \bibinfo {author} {\bibfnamefont {F.}~\bibnamefont {Mila}}, \ and\ \bibinfo
  {author} {\bibfnamefont {A.~M.}\ \bibnamefont {L\"auchli}},\ }\href {\doibase
  10.1103/PhysRevB.86.041106} {\bibfield  {journal} {\bibinfo  {journal} {Phys.
  Rev. B}\ }\textbf {\bibinfo {volume} {86}},\ \bibinfo {pages} {041106}
  (\bibinfo {year} {2012})}\BibitemShut {NoStop}%
\bibitem [{\citenamefont {Corboz}\ \emph {et~al.}(2013)\citenamefont {Corboz},
  \citenamefont {Lajk\'o}, \citenamefont {Penc}, \citenamefont {Mila},\ and\
  \citenamefont {L\"auchli}}]{Corboz_2013_PRB}%
  \BibitemOpen
  \bibfield  {author} {\bibinfo {author} {\bibfnamefont {P.}~\bibnamefont
  {Corboz}}, \bibinfo {author} {\bibfnamefont {M.}~\bibnamefont {Lajk\'o}},
  \bibinfo {author} {\bibfnamefont {K.}~\bibnamefont {Penc}}, \bibinfo {author}
  {\bibfnamefont {F.}~\bibnamefont {Mila}}, \ and\ \bibinfo {author}
  {\bibfnamefont {A.~M.}\ \bibnamefont {L\"auchli}},\ }\href {\doibase
  10.1103/PhysRevB.87.195113} {\bibfield  {journal} {\bibinfo  {journal} {Phys.
  Rev. B}\ }\textbf {\bibinfo {volume} {87}},\ \bibinfo {pages} {195113}
  (\bibinfo {year} {2013})}\BibitemShut {NoStop}%
\bibitem [{\citenamefont {Nataf}\ \emph {et~al.}(2016)\citenamefont {Nataf},
  \citenamefont {Lajk\'o}, \citenamefont {Corboz}, \citenamefont {L\"auchli},
  \citenamefont {Penc},\ and\ \citenamefont {Mila}}]{Nataf_2016_PRB}%
  \BibitemOpen
  \bibfield  {author} {\bibinfo {author} {\bibfnamefont {P.}~\bibnamefont
  {Nataf}}, \bibinfo {author} {\bibfnamefont {M.}~\bibnamefont {Lajk\'o}},
  \bibinfo {author} {\bibfnamefont {P.}~\bibnamefont {Corboz}}, \bibinfo
  {author} {\bibfnamefont {A.~M.}\ \bibnamefont {L\"auchli}}, \bibinfo {author}
  {\bibfnamefont {K.}~\bibnamefont {Penc}}, \ and\ \bibinfo {author}
  {\bibfnamefont {F.}~\bibnamefont {Mila}},\ }\href {\doibase
  10.1103/PhysRevB.93.201113} {\bibfield  {journal} {\bibinfo  {journal} {Phys.
  Rev. B}\ }\textbf {\bibinfo {volume} {93}},\ \bibinfo {pages} {201113}
  (\bibinfo {year} {2016})}\BibitemShut {NoStop}%
\bibitem [{\citenamefont {Corboz}\ \emph {et~al.}(2011)\citenamefont {Corboz},
  \citenamefont {L\"auchli}, \citenamefont {Penc}, \citenamefont {Troyer},\
  and\ \citenamefont {Mila}}]{Corboz_2011_PRL}%
  \BibitemOpen
  \bibfield  {author} {\bibinfo {author} {\bibfnamefont {P.}~\bibnamefont
  {Corboz}}, \bibinfo {author} {\bibfnamefont {A.~M.}\ \bibnamefont
  {L\"auchli}}, \bibinfo {author} {\bibfnamefont {K.}~\bibnamefont {Penc}},
  \bibinfo {author} {\bibfnamefont {M.}~\bibnamefont {Troyer}}, \ and\ \bibinfo
  {author} {\bibfnamefont {F.}~\bibnamefont {Mila}},\ }\href {\doibase
  10.1103/PhysRevLett.107.215301} {\bibfield  {journal} {\bibinfo  {journal}
  {Phys. Rev. Lett.}\ }\textbf {\bibinfo {volume} {107}},\ \bibinfo {pages}
  {215301} (\bibinfo {year} {2011})}\BibitemShut {NoStop}%
\bibitem [{\citenamefont {Bauer}\ \emph {et~al.}(2012)\citenamefont {Bauer},
  \citenamefont {Corboz}, \citenamefont {L\"auchli}, \citenamefont {Messio},
  \citenamefont {Penc}, \citenamefont {Troyer},\ and\ \citenamefont
  {Mila}}]{Bauer_2012_PRB}%
  \BibitemOpen
  \bibfield  {author} {\bibinfo {author} {\bibfnamefont {B.}~\bibnamefont
  {Bauer}}, \bibinfo {author} {\bibfnamefont {P.}~\bibnamefont {Corboz}},
  \bibinfo {author} {\bibfnamefont {A.~M.}\ \bibnamefont {L\"auchli}}, \bibinfo
  {author} {\bibfnamefont {L.}~\bibnamefont {Messio}}, \bibinfo {author}
  {\bibfnamefont {K.}~\bibnamefont {Penc}}, \bibinfo {author} {\bibfnamefont
  {M.}~\bibnamefont {Troyer}}, \ and\ \bibinfo {author} {\bibfnamefont
  {F.}~\bibnamefont {Mila}},\ }\href {\doibase 10.1103/PhysRevB.85.125116}
  {\bibfield  {journal} {\bibinfo  {journal} {Phys. Rev. B}\ }\textbf {\bibinfo
  {volume} {85}},\ \bibinfo {pages} {125116} (\bibinfo {year}
  {2012})}\BibitemShut {NoStop}%
\bibitem [{\citenamefont {Weichselbaum}\ \emph {et~al.}(2018)\citenamefont
  {Weichselbaum}, \citenamefont {Capponi}, \citenamefont {Lecheminant},
  \citenamefont {Tsvelik},\ and\ \citenamefont
  {L{\"a}uchli}}]{weichselbaum2018unified}%
  \BibitemOpen
  \bibfield  {author} {\bibinfo {author} {\bibfnamefont {A.}~\bibnamefont
  {Weichselbaum}}, \bibinfo {author} {\bibfnamefont {S.}~\bibnamefont
  {Capponi}}, \bibinfo {author} {\bibfnamefont {P.}~\bibnamefont
  {Lecheminant}}, \bibinfo {author} {\bibfnamefont {A.~M.}\ \bibnamefont
  {Tsvelik}}, \ and\ \bibinfo {author} {\bibfnamefont {A.~M.}\ \bibnamefont
  {L{\"a}uchli}},\ }\href@noop {} {\bibfield  {journal} {\bibinfo  {journal}
  {Physical Review B}\ }\textbf {\bibinfo {volume} {98}},\ \bibinfo {pages}
  {085104} (\bibinfo {year} {2018})}\BibitemShut {NoStop}%
\bibitem [{\citenamefont {Richardson}(1997)}]{Richardson_1997_RMP}%
  \BibitemOpen
  \bibfield  {author} {\bibinfo {author} {\bibfnamefont {R.~C.}\ \bibnamefont
  {Richardson}},\ }\href {\doibase 10.1103/RevModPhys.69.683} {\bibfield
  {journal} {\bibinfo  {journal} {Rev. Mod. Phys.}\ }\textbf {\bibinfo {volume}
  {69}},\ \bibinfo {pages} {683} (\bibinfo {year} {1997})}\BibitemShut
  {NoStop}%
\bibitem [{\citenamefont {Bernier}\ \emph {et~al.}(2009)\citenamefont
  {Bernier}, \citenamefont {Kollath}, \citenamefont {Georges}, \citenamefont
  {De~Leo}, \citenamefont {Gerbier}, \citenamefont {Salomon},\ and\
  \citenamefont {K\"ohl}}]{Bernier_2009_PRA}%
  \BibitemOpen
  \bibfield  {author} {\bibinfo {author} {\bibfnamefont {J.-S.}\ \bibnamefont
  {Bernier}}, \bibinfo {author} {\bibfnamefont {C.}~\bibnamefont {Kollath}},
  \bibinfo {author} {\bibfnamefont {A.}~\bibnamefont {Georges}}, \bibinfo
  {author} {\bibfnamefont {L.}~\bibnamefont {De~Leo}}, \bibinfo {author}
  {\bibfnamefont {F.}~\bibnamefont {Gerbier}}, \bibinfo {author} {\bibfnamefont
  {C.}~\bibnamefont {Salomon}}, \ and\ \bibinfo {author} {\bibfnamefont
  {M.}~\bibnamefont {K\"ohl}},\ }\href {\doibase 10.1103/PhysRevA.79.061601}
  {\bibfield  {journal} {\bibinfo  {journal} {Phys. Rev. A}\ }\textbf {\bibinfo
  {volume} {79}},\ \bibinfo {pages} {061601} (\bibinfo {year}
  {2009})}\BibitemShut {NoStop}%
\bibitem [{\citenamefont {Bonnes}\ \emph {et~al.}(2012)\citenamefont {Bonnes},
  \citenamefont {Hazzard}, \citenamefont {Manmana}, \citenamefont {Rey},\ and\
  \citenamefont {Wessel}}]{Bonnes_2012_PRL}%
  \BibitemOpen
  \bibfield  {author} {\bibinfo {author} {\bibfnamefont {L.}~\bibnamefont
  {Bonnes}}, \bibinfo {author} {\bibfnamefont {K.~R.~A.}\ \bibnamefont
  {Hazzard}}, \bibinfo {author} {\bibfnamefont {S.~R.}\ \bibnamefont
  {Manmana}}, \bibinfo {author} {\bibfnamefont {A.~M.}\ \bibnamefont {Rey}}, \
  and\ \bibinfo {author} {\bibfnamefont {S.}~\bibnamefont {Wessel}},\ }\href
  {\doibase 10.1103/PhysRevLett.109.205305} {\bibfield  {journal} {\bibinfo
  {journal} {Phys. Rev. Lett.}\ }\textbf {\bibinfo {volume} {109}},\ \bibinfo
  {pages} {205305} (\bibinfo {year} {2012})}\BibitemShut {NoStop}%
\bibitem [{\citenamefont {Hazzard}\ \emph {et~al.}(2012)\citenamefont
  {Hazzard}, \citenamefont {Gurarie}, \citenamefont {Hermele},\ and\
  \citenamefont {Rey}}]{Hazzard_2012_PRA}%
  \BibitemOpen
  \bibfield  {author} {\bibinfo {author} {\bibfnamefont {K.~R.~A.}\
  \bibnamefont {Hazzard}}, \bibinfo {author} {\bibfnamefont {V.}~\bibnamefont
  {Gurarie}}, \bibinfo {author} {\bibfnamefont {M.}~\bibnamefont {Hermele}}, \
  and\ \bibinfo {author} {\bibfnamefont {A.~M.}\ \bibnamefont {Rey}},\ }\href
  {\doibase 10.1103/PhysRevA.85.041604} {\bibfield  {journal} {\bibinfo
  {journal} {Phys. Rev. A}\ }\textbf {\bibinfo {volume} {85}},\ \bibinfo
  {pages} {041604} (\bibinfo {year} {2012})}\BibitemShut {NoStop}%
\bibitem [{\citenamefont {Werner}\ \emph {et~al.}(2005)\citenamefont {Werner},
  \citenamefont {Parcollet}, \citenamefont {Georges},\ and\ \citenamefont
  {Hassan}}]{Werner_2005_PRL}%
  \BibitemOpen
  \bibfield  {author} {\bibinfo {author} {\bibfnamefont {F.}~\bibnamefont
  {Werner}}, \bibinfo {author} {\bibfnamefont {O.}~\bibnamefont {Parcollet}},
  \bibinfo {author} {\bibfnamefont {A.}~\bibnamefont {Georges}}, \ and\
  \bibinfo {author} {\bibfnamefont {S.~R.}\ \bibnamefont {Hassan}},\ }\href
  {\doibase 10.1103/PhysRevLett.95.056401} {\bibfield  {journal} {\bibinfo
  {journal} {Phys. Rev. Lett.}\ }\textbf {\bibinfo {volume} {95}},\ \bibinfo
  {pages} {056401} (\bibinfo {year} {2005})}\BibitemShut {NoStop}%
\bibitem [{\citenamefont {Blakie}\ and\ \citenamefont
  {Bezett}(2005)}]{Blakie_2005_PRA}%
  \BibitemOpen
  \bibfield  {author} {\bibinfo {author} {\bibfnamefont {P.~B.}\ \bibnamefont
  {Blakie}}\ and\ \bibinfo {author} {\bibfnamefont {A.}~\bibnamefont
  {Bezett}},\ }\href {\doibase 10.1103/PhysRevA.71.033616} {\bibfield
  {journal} {\bibinfo  {journal} {Phys. Rev. A}\ }\textbf {\bibinfo {volume}
  {71}},\ \bibinfo {pages} {033616} (\bibinfo {year} {2005})}\BibitemShut
  {NoStop}%
\bibitem [{\citenamefont {Blakie}\ \emph {et~al.}(2007)\citenamefont {Blakie},
  \citenamefont {Bezett},\ and\ \citenamefont {Buonsante}}]{Blakie_2007_PRA}%
  \BibitemOpen
  \bibfield  {author} {\bibinfo {author} {\bibfnamefont {P.~B.}\ \bibnamefont
  {Blakie}}, \bibinfo {author} {\bibfnamefont {A.}~\bibnamefont {Bezett}}, \
  and\ \bibinfo {author} {\bibfnamefont {P.}~\bibnamefont {Buonsante}},\ }\href
  {\doibase 10.1103/PhysRevA.75.063609} {\bibfield  {journal} {\bibinfo
  {journal} {Phys. Rev. A}\ }\textbf {\bibinfo {volume} {75}},\ \bibinfo
  {pages} {063609} (\bibinfo {year} {2007})}\BibitemShut {NoStop}%
\bibitem [{\citenamefont {Taie}\ \emph {et~al.}(2012)\citenamefont {Taie},
  \citenamefont {Yamazaki}, \citenamefont {Sugawa},\ and\ \citenamefont
  {Takahashi}}]{Taie_2012_NatPhys}%
  \BibitemOpen
  \bibfield  {author} {\bibinfo {author} {\bibfnamefont {S.}~\bibnamefont
  {Taie}}, \bibinfo {author} {\bibfnamefont {R.}~\bibnamefont {Yamazaki}},
  \bibinfo {author} {\bibfnamefont {S.}~\bibnamefont {Sugawa}}, \ and\ \bibinfo
  {author} {\bibfnamefont {Y.}~\bibnamefont {Takahashi}},\ }\href@noop {}
  {\bibfield  {journal} {\bibinfo  {journal} {Nature Physics}\ }\textbf
  {\bibinfo {volume} {8}},\ \bibinfo {pages} {825} (\bibinfo {year}
  {2012})}\BibitemShut {NoStop}%
\bibitem [{\citenamefont {Mazurenko}\ \emph {et~al.}(2017)\citenamefont
  {Mazurenko}, \citenamefont {Chiu}, \citenamefont {Ji}, \citenamefont
  {Parsons}, \citenamefont {Kanász-Nagy}, \citenamefont {Schmidt},
  \citenamefont {Grusdt}, \citenamefont {Demler}, \citenamefont {Greif},\ and\
  \citenamefont {Greiner}}]{Mazurenko2017}%
  \BibitemOpen
  \bibfield  {author} {\bibinfo {author} {\bibfnamefont {A.}~\bibnamefont
  {Mazurenko}}, \bibinfo {author} {\bibfnamefont {C.~S.}\ \bibnamefont {Chiu}},
  \bibinfo {author} {\bibfnamefont {G.}~\bibnamefont {Ji}}, \bibinfo {author}
  {\bibfnamefont {M.~F.}\ \bibnamefont {Parsons}}, \bibinfo {author}
  {\bibfnamefont {M.}~\bibnamefont {Kanász-Nagy}}, \bibinfo {author}
  {\bibfnamefont {R.}~\bibnamefont {Schmidt}}, \bibinfo {author} {\bibfnamefont
  {F.}~\bibnamefont {Grusdt}}, \bibinfo {author} {\bibfnamefont
  {E.}~\bibnamefont {Demler}}, \bibinfo {author} {\bibfnamefont
  {D.}~\bibnamefont {Greif}}, \ and\ \bibinfo {author} {\bibfnamefont
  {M.}~\bibnamefont {Greiner}},\ }\href {https://doi.org/10.1038/nature22362}
  {\bibfield  {journal} {\bibinfo  {journal} {Nature}\ }\textbf {\bibinfo
  {volume} {545}},\ \bibinfo {pages} {462 EP } (\bibinfo {year}
  {2017})}\BibitemShut {NoStop}%
\bibitem [{\citenamefont {Chiu}\ \emph {et~al.}(2018)\citenamefont {Chiu},
  \citenamefont {Ji}, \citenamefont {Mazurenko}, \citenamefont {Greif},\ and\
  \citenamefont {Greiner}}]{Chiu_2018_PRL}%
  \BibitemOpen
  \bibfield  {author} {\bibinfo {author} {\bibfnamefont {C.~S.}\ \bibnamefont
  {Chiu}}, \bibinfo {author} {\bibfnamefont {G.}~\bibnamefont {Ji}}, \bibinfo
  {author} {\bibfnamefont {A.}~\bibnamefont {Mazurenko}}, \bibinfo {author}
  {\bibfnamefont {D.}~\bibnamefont {Greif}}, \ and\ \bibinfo {author}
  {\bibfnamefont {M.}~\bibnamefont {Greiner}},\ }\href {\doibase
  10.1103/PhysRevLett.120.243201} {\bibfield  {journal} {\bibinfo  {journal}
  {Phys. Rev. Lett.}\ }\textbf {\bibinfo {volume} {120}},\ \bibinfo {pages}
  {243201} (\bibinfo {year} {2018})}\BibitemShut {NoStop}%
\bibitem [{\citenamefont {Greif}\ \emph {et~al.}(2013)\citenamefont {Greif},
  \citenamefont {Uehlinger}, \citenamefont {Jotzu}, \citenamefont {Tarruell},\
  and\ \citenamefont {Esslinger}}]{Greif_2013_Science}%
  \BibitemOpen
  \bibfield  {author} {\bibinfo {author} {\bibfnamefont {D.}~\bibnamefont
  {Greif}}, \bibinfo {author} {\bibfnamefont {T.}~\bibnamefont {Uehlinger}},
  \bibinfo {author} {\bibfnamefont {G.}~\bibnamefont {Jotzu}}, \bibinfo
  {author} {\bibfnamefont {L.}~\bibnamefont {Tarruell}}, \ and\ \bibinfo
  {author} {\bibfnamefont {T.}~\bibnamefont {Esslinger}},\ }\href@noop {}
  {\bibfield  {journal} {\bibinfo  {journal} {Science}\ }\textbf {\bibinfo
  {volume} {340}},\ \bibinfo {pages} {1307} (\bibinfo {year}
  {2013})}\BibitemShut {NoStop}%
\bibitem [{\citenamefont {Gaunt}\ \emph {et~al.}(2013)\citenamefont {Gaunt},
  \citenamefont {Schmidutz}, \citenamefont {Gotlibovych}, \citenamefont
  {Smith},\ and\ \citenamefont {Hadzibabic}}]{Gaunt_2013_PRL}%
  \BibitemOpen
  \bibfield  {author} {\bibinfo {author} {\bibfnamefont {A.~L.}\ \bibnamefont
  {Gaunt}}, \bibinfo {author} {\bibfnamefont {T.~F.}\ \bibnamefont
  {Schmidutz}}, \bibinfo {author} {\bibfnamefont {I.}~\bibnamefont
  {Gotlibovych}}, \bibinfo {author} {\bibfnamefont {R.~P.}\ \bibnamefont
  {Smith}}, \ and\ \bibinfo {author} {\bibfnamefont {Z.}~\bibnamefont
  {Hadzibabic}},\ }\href {\doibase 10.1103/PhysRevLett.110.200406} {\bibfield
  {journal} {\bibinfo  {journal} {Phys. Rev. Lett.}\ }\textbf {\bibinfo
  {volume} {110}},\ \bibinfo {pages} {200406} (\bibinfo {year}
  {2013})}\BibitemShut {NoStop}%
\bibitem [{\citenamefont {Oitmaa}\ \emph {et~al.}(2006)\citenamefont {Oitmaa},
  \citenamefont {Hamer},\ and\ \citenamefont {Zheng}}]{oitmaa}%
  \BibitemOpen
  \bibfield  {author} {\bibinfo {author} {\bibfnamefont {J.}~\bibnamefont
  {Oitmaa}}, \bibinfo {author} {\bibfnamefont {C.}~\bibnamefont {Hamer}}, \
  and\ \bibinfo {author} {\bibfnamefont {W.}~\bibnamefont {Zheng}},\ }\href
  {\doibase 10.1017/CBO9780511584398} {\emph {\bibinfo {title} {Series
  Expansion Methods for Strongly Interacting Lattice Models}}}\ (\bibinfo
  {publisher} {Cambridge University Press},\ \bibinfo {year}
  {2006})\BibitemShut {NoStop}%
\bibitem [{Note1()}]{Note1}%
  \BibitemOpen
  \bibinfo {note} {In this context Ref. \cite {Bernier_2009_PRA} discusses the
  improvement in cooling when flattening the harmonic profile of the reservoir,
  resulting in flat (box-like) profile considered here.}\BibitemShut {Stop}%
\bibitem [{Note2()}]{Note2}%
  \BibitemOpen
  \bibinfo {note} {See e.g. \cite {Bonnes_2012_PRL} or Fig. 1 in \cite
  {Messio_2012_PRL}, which analyzed the entropy density for a one-dimensional
  chain with $c_\ell =2$. Since we rely on LDA, we expect the dependence of
  $\protect \bar {s}_D$ to qualitatively hold for the square lattice with
  $c_\ell =4$ as it appears only as a prefactor in Eq.~(\ref
  {TwoVarietySecondOrderTerm}).}\BibitemShut {Stop}%
\bibitem [{\citenamefont {J\"ordens}\ \emph {et~al.}(2010)\citenamefont
  {J\"ordens}, \citenamefont {Tarruell}, \citenamefont {Greif}, \citenamefont
  {Uehlinger}, \citenamefont {Strohmaier}, \citenamefont {Moritz},
  \citenamefont {Esslinger}, \citenamefont {De~Leo}, \citenamefont {Kollath},
  \citenamefont {Georges}, \citenamefont {Scarola}, \citenamefont {Pollet},
  \citenamefont {Burovski}, \citenamefont {Kozik},\ and\ \citenamefont
  {Troyer}}]{Jordens_2010_PRL}%
  \BibitemOpen
  \bibfield  {author} {\bibinfo {author} {\bibfnamefont {R.}~\bibnamefont
  {J\"ordens}}, \bibinfo {author} {\bibfnamefont {L.}~\bibnamefont {Tarruell}},
  \bibinfo {author} {\bibfnamefont {D.}~\bibnamefont {Greif}}, \bibinfo
  {author} {\bibfnamefont {T.}~\bibnamefont {Uehlinger}}, \bibinfo {author}
  {\bibfnamefont {N.}~\bibnamefont {Strohmaier}}, \bibinfo {author}
  {\bibfnamefont {H.}~\bibnamefont {Moritz}}, \bibinfo {author} {\bibfnamefont
  {T.}~\bibnamefont {Esslinger}}, \bibinfo {author} {\bibfnamefont
  {L.}~\bibnamefont {De~Leo}}, \bibinfo {author} {\bibfnamefont
  {C.}~\bibnamefont {Kollath}}, \bibinfo {author} {\bibfnamefont
  {A.}~\bibnamefont {Georges}}, \bibinfo {author} {\bibfnamefont
  {V.}~\bibnamefont {Scarola}}, \bibinfo {author} {\bibfnamefont
  {L.}~\bibnamefont {Pollet}}, \bibinfo {author} {\bibfnamefont
  {E.}~\bibnamefont {Burovski}}, \bibinfo {author} {\bibfnamefont
  {E.}~\bibnamefont {Kozik}}, \ and\ \bibinfo {author} {\bibfnamefont
  {M.}~\bibnamefont {Troyer}},\ }\href {\doibase
  10.1103/PhysRevLett.104.180401} {\bibfield  {journal} {\bibinfo  {journal}
  {Phys. Rev. Lett.}\ }\textbf {\bibinfo {volume} {104}},\ \bibinfo {pages}
  {180401} (\bibinfo {year} {2010})}\BibitemShut {NoStop}%
\bibitem [{\citenamefont {LeBlanc}\ \emph {et~al.}(2015)\citenamefont
  {LeBlanc}, \citenamefont {Antipov}, \citenamefont {Becca}, \citenamefont
  {Bulik}, \citenamefont {Chan}, \citenamefont {Chung}, \citenamefont {Deng},
  \citenamefont {Ferrero}, \citenamefont {Henderson}, \citenamefont
  {Jim\'enez-Hoyos}, \citenamefont {Kozik}, \citenamefont {Liu}, \citenamefont
  {Millis}, \citenamefont {Prokof'ev}, \citenamefont {Qin}, \citenamefont
  {Scuseria}, \citenamefont {Shi}, \citenamefont {Svistunov}, \citenamefont
  {Tocchio}, \citenamefont {Tupitsyn}, \citenamefont {White}, \citenamefont
  {Zhang}, \citenamefont {Zheng}, \citenamefont {Zhu},\ and\ \citenamefont
  {Gull}}]{PhysRevX.5.041041}%
  \BibitemOpen
  \bibfield  {author} {\bibinfo {author} {\bibfnamefont {J.~P.~F.}\
  \bibnamefont {LeBlanc}}, \bibinfo {author} {\bibfnamefont {A.~E.}\
  \bibnamefont {Antipov}}, \bibinfo {author} {\bibfnamefont {F.}~\bibnamefont
  {Becca}}, \bibinfo {author} {\bibfnamefont {I.~W.}\ \bibnamefont {Bulik}},
  \bibinfo {author} {\bibfnamefont {G.~K.-L.}\ \bibnamefont {Chan}}, \bibinfo
  {author} {\bibfnamefont {C.-M.}\ \bibnamefont {Chung}}, \bibinfo {author}
  {\bibfnamefont {Y.}~\bibnamefont {Deng}}, \bibinfo {author} {\bibfnamefont
  {M.}~\bibnamefont {Ferrero}}, \bibinfo {author} {\bibfnamefont {T.~M.}\
  \bibnamefont {Henderson}}, \bibinfo {author} {\bibfnamefont {C.~A.}\
  \bibnamefont {Jim\'enez-Hoyos}}, \bibinfo {author} {\bibfnamefont
  {E.}~\bibnamefont {Kozik}}, \bibinfo {author} {\bibfnamefont {X.-W.}\
  \bibnamefont {Liu}}, \bibinfo {author} {\bibfnamefont {A.~J.}\ \bibnamefont
  {Millis}}, \bibinfo {author} {\bibfnamefont {N.~V.}\ \bibnamefont
  {Prokof'ev}}, \bibinfo {author} {\bibfnamefont {M.}~\bibnamefont {Qin}},
  \bibinfo {author} {\bibfnamefont {G.~E.}\ \bibnamefont {Scuseria}}, \bibinfo
  {author} {\bibfnamefont {H.}~\bibnamefont {Shi}}, \bibinfo {author}
  {\bibfnamefont {B.~V.}\ \bibnamefont {Svistunov}}, \bibinfo {author}
  {\bibfnamefont {L.~F.}\ \bibnamefont {Tocchio}}, \bibinfo {author}
  {\bibfnamefont {I.~S.}\ \bibnamefont {Tupitsyn}}, \bibinfo {author}
  {\bibfnamefont {S.~R.}\ \bibnamefont {White}}, \bibinfo {author}
  {\bibfnamefont {S.}~\bibnamefont {Zhang}}, \bibinfo {author} {\bibfnamefont
  {B.-X.}\ \bibnamefont {Zheng}}, \bibinfo {author} {\bibfnamefont
  {Z.}~\bibnamefont {Zhu}}, \ and\ \bibinfo {author} {\bibfnamefont
  {E.}~\bibnamefont {Gull}} (\bibinfo {collaboration} {Simons Collaboration on
  the Many-Electron Problem}),\ }\href {\doibase 10.1103/PhysRevX.5.041041}
  {\bibfield  {journal} {\bibinfo  {journal} {Phys. Rev. X}\ }\textbf {\bibinfo
  {volume} {5}},\ \bibinfo {pages} {041041} (\bibinfo {year}
  {2015})}\BibitemShut {NoStop}%
\bibitem [{\citenamefont {Stellmer}\ \emph {et~al.}(2011)\citenamefont
  {Stellmer}, \citenamefont {Grimm},\ and\ \citenamefont
  {Schreck}}]{Stellmer_2011_PRA}%
  \BibitemOpen
  \bibfield  {author} {\bibinfo {author} {\bibfnamefont {S.}~\bibnamefont
  {Stellmer}}, \bibinfo {author} {\bibfnamefont {R.}~\bibnamefont {Grimm}}, \
  and\ \bibinfo {author} {\bibfnamefont {F.}~\bibnamefont {Schreck}},\ }\href
  {\doibase 10.1103/PhysRevA.84.043611} {\bibfield  {journal} {\bibinfo
  {journal} {Phys. Rev. A}\ }\textbf {\bibinfo {volume} {84}},\ \bibinfo
  {pages} {043611} (\bibinfo {year} {2011})}\BibitemShut {NoStop}%
\bibitem [{\citenamefont {Grimm}\ \emph {et~al.}(2000)\citenamefont {Grimm},
  \citenamefont {Weidem{\"u}ller},\ and\ \citenamefont
  {Ovchinnikov}}]{grimm2000optical}%
  \BibitemOpen
  \bibfield  {author} {\bibinfo {author} {\bibfnamefont {R.}~\bibnamefont
  {Grimm}}, \bibinfo {author} {\bibfnamefont {M.}~\bibnamefont
  {Weidem{\"u}ller}}, \ and\ \bibinfo {author} {\bibfnamefont {Y.~B.}\
  \bibnamefont {Ovchinnikov}},\ }\href@noop {} {\bibfield  {journal} {\bibinfo
  {journal} {Advances in atomic, molecular, and optical physics}\ }\textbf
  {\bibinfo {volume} {42}},\ \bibinfo {pages} {95} (\bibinfo {year}
  {2000})}\BibitemShut {NoStop}%
\bibitem [{\citenamefont {Onishchenko}\ \emph {et~al.}(2019)\citenamefont
  {Onishchenko}, \citenamefont {Pyatchenkov}, \citenamefont {Urech},
  \citenamefont {Chen}, \citenamefont {Bennetts}, \citenamefont {Siviloglou},\
  and\ \citenamefont {Schreck}}]{Onishchenko_2019_PRA}%
  \BibitemOpen
  \bibfield  {author} {\bibinfo {author} {\bibfnamefont {O.}~\bibnamefont
  {Onishchenko}}, \bibinfo {author} {\bibfnamefont {S.}~\bibnamefont
  {Pyatchenkov}}, \bibinfo {author} {\bibfnamefont {A.}~\bibnamefont {Urech}},
  \bibinfo {author} {\bibfnamefont {C.-C.}\ \bibnamefont {Chen}}, \bibinfo
  {author} {\bibfnamefont {S.}~\bibnamefont {Bennetts}}, \bibinfo {author}
  {\bibfnamefont {G.~A.}\ \bibnamefont {Siviloglou}}, \ and\ \bibinfo {author}
  {\bibfnamefont {F.}~\bibnamefont {Schreck}},\ }\href {\doibase
  10.1103/PhysRevA.99.052503} {\bibfield  {journal} {\bibinfo  {journal} {Phys.
  Rev. A}\ }\textbf {\bibinfo {volume} {99}},\ \bibinfo {pages} {052503}
  (\bibinfo {year} {2019})}\BibitemShut {NoStop}%
\bibitem [{Note3()}]{Note3}%
  \BibitemOpen
  \bibinfo {note} {Alex Urech, private communication}\BibitemShut {NoStop}%
\bibitem [{\citenamefont {Boyd}\ \emph {et~al.}(2007)\citenamefont {Boyd},
  \citenamefont {Zelevinsky}, \citenamefont {Ludlow}, \citenamefont {Blatt},
  \citenamefont {Zanon-Willette}, \citenamefont {Foreman},\ and\ \citenamefont
  {Ye}}]{Boyd_2007_PRA}%
  \BibitemOpen
  \bibfield  {author} {\bibinfo {author} {\bibfnamefont {M.~M.}\ \bibnamefont
  {Boyd}}, \bibinfo {author} {\bibfnamefont {T.}~\bibnamefont {Zelevinsky}},
  \bibinfo {author} {\bibfnamefont {A.~D.}\ \bibnamefont {Ludlow}}, \bibinfo
  {author} {\bibfnamefont {S.}~\bibnamefont {Blatt}}, \bibinfo {author}
  {\bibfnamefont {T.}~\bibnamefont {Zanon-Willette}}, \bibinfo {author}
  {\bibfnamefont {S.~M.}\ \bibnamefont {Foreman}}, \ and\ \bibinfo {author}
  {\bibfnamefont {J.}~\bibnamefont {Ye}},\ }\href {\doibase
  10.1103/PhysRevA.76.022510} {\bibfield  {journal} {\bibinfo  {journal} {Phys.
  Rev. A}\ }\textbf {\bibinfo {volume} {76}},\ \bibinfo {pages} {022510}
  (\bibinfo {year} {2007})}\BibitemShut {NoStop}%
\bibitem [{\citenamefont {Berends}\ and\ \citenamefont
  {Maleki}(1992)}]{Berends_1992_JOSA}%
  \BibitemOpen
  \bibfield  {author} {\bibinfo {author} {\bibfnamefont {R.}~\bibnamefont
  {Berends}}\ and\ \bibinfo {author} {\bibfnamefont {L.}~\bibnamefont
  {Maleki}},\ }\href@noop {} {\bibfield  {journal} {\bibinfo  {journal} {JOSA
  B}\ }\textbf {\bibinfo {volume} {9}},\ \bibinfo {pages} {332} (\bibinfo
  {year} {1992})}\BibitemShut {NoStop}%
\bibitem [{\citenamefont {Shibata}\ \emph {et~al.}(2014)\citenamefont
  {Shibata}, \citenamefont {Yamamoto}, \citenamefont {Seki},\ and\
  \citenamefont {Takahashi}}]{Shibata_2014_PRA}%
  \BibitemOpen
  \bibfield  {author} {\bibinfo {author} {\bibfnamefont {K.}~\bibnamefont
  {Shibata}}, \bibinfo {author} {\bibfnamefont {R.}~\bibnamefont {Yamamoto}},
  \bibinfo {author} {\bibfnamefont {Y.}~\bibnamefont {Seki}}, \ and\ \bibinfo
  {author} {\bibfnamefont {Y.}~\bibnamefont {Takahashi}},\ }\href {\doibase
  10.1103/PhysRevA.89.031601} {\bibfield  {journal} {\bibinfo  {journal} {Phys.
  Rev. A}\ }\textbf {\bibinfo {volume} {89}},\ \bibinfo {pages} {031601}
  (\bibinfo {year} {2014})}\BibitemShut {NoStop}%
\bibitem [{\citenamefont {Ludlow}\ \emph {et~al.}(2015)\citenamefont {Ludlow},
  \citenamefont {Boyd}, \citenamefont {Ye}, \citenamefont {Peik},\ and\
  \citenamefont {Schmidt}}]{Ludlow_2015_RMP}%
  \BibitemOpen
  \bibfield  {author} {\bibinfo {author} {\bibfnamefont {A.~D.}\ \bibnamefont
  {Ludlow}}, \bibinfo {author} {\bibfnamefont {M.~M.}\ \bibnamefont {Boyd}},
  \bibinfo {author} {\bibfnamefont {J.}~\bibnamefont {Ye}}, \bibinfo {author}
  {\bibfnamefont {E.}~\bibnamefont {Peik}}, \ and\ \bibinfo {author}
  {\bibfnamefont {P.~O.}\ \bibnamefont {Schmidt}},\ }\href {\doibase
  10.1103/RevModPhys.87.637} {\bibfield  {journal} {\bibinfo  {journal} {Rev.
  Mod. Phys.}\ }\textbf {\bibinfo {volume} {87}},\ \bibinfo {pages} {637}
  (\bibinfo {year} {2015})}\BibitemShut {NoStop}%
\bibitem [{\citenamefont {Takasu}\ \emph {et~al.}(2017)\citenamefont {Takasu},
  \citenamefont {Fukushima}, \citenamefont {Nakamura},\ and\ \citenamefont
  {Takahashi}}]{Takasu_2017_PRA}%
  \BibitemOpen
  \bibfield  {author} {\bibinfo {author} {\bibfnamefont {Y.}~\bibnamefont
  {Takasu}}, \bibinfo {author} {\bibfnamefont {Y.}~\bibnamefont {Fukushima}},
  \bibinfo {author} {\bibfnamefont {Y.}~\bibnamefont {Nakamura}}, \ and\
  \bibinfo {author} {\bibfnamefont {Y.}~\bibnamefont {Takahashi}},\ }\href
  {\doibase 10.1103/PhysRevA.96.023602} {\bibfield  {journal} {\bibinfo
  {journal} {Phys. Rev. A}\ }\textbf {\bibinfo {volume} {96}},\ \bibinfo
  {pages} {023602} (\bibinfo {year} {2017})}\BibitemShut {NoStop}%
\bibitem [{\citenamefont {Messio}\ and\ \citenamefont
  {Mila}(2012)}]{Messio_2012_PRL}%
  \BibitemOpen
  \bibfield  {author} {\bibinfo {author} {\bibfnamefont {L.}~\bibnamefont
  {Messio}}\ and\ \bibinfo {author} {\bibfnamefont {F.}~\bibnamefont {Mila}},\
  }\href {\doibase 10.1103/PhysRevLett.109.205306} {\bibfield  {journal}
  {\bibinfo  {journal} {Phys. Rev. Lett.}\ }\textbf {\bibinfo {volume} {109}},\
  \bibinfo {pages} {205306} (\bibinfo {year} {2012})}\BibitemShut {NoStop}%
\bibitem [{\citenamefont {Henderson}\ \emph {et~al.}(1992)\citenamefont
  {Henderson}, \citenamefont {Oitmaa},\ and\ \citenamefont
  {Ashley}}]{Henderson_1992_PRB}%
  \BibitemOpen
  \bibfield  {author} {\bibinfo {author} {\bibfnamefont {J.~A.}\ \bibnamefont
  {Henderson}}, \bibinfo {author} {\bibfnamefont {J.}~\bibnamefont {Oitmaa}}, \
  and\ \bibinfo {author} {\bibfnamefont {M.~C.~B.}\ \bibnamefont {Ashley}},\
  }\href {\doibase 10.1103/PhysRevB.46.6328} {\bibfield  {journal} {\bibinfo
  {journal} {Phys. Rev. B}\ }\textbf {\bibinfo {volume} {46}},\ \bibinfo
  {pages} {6328} (\bibinfo {year} {1992})}\BibitemShut {NoStop}%
\bibitem [{\citenamefont {Pan}\ and\ \citenamefont
  {Wang}(1991{\natexlab{a}})}]{Pan_1991_PRB}%
  \BibitemOpen
  \bibfield  {author} {\bibinfo {author} {\bibfnamefont {K.-K.}\ \bibnamefont
  {Pan}}\ and\ \bibinfo {author} {\bibfnamefont {Y.-L.}\ \bibnamefont {Wang}},\
  }\href {\doibase 10.1103/PhysRevB.43.3706} {\bibfield  {journal} {\bibinfo
  {journal} {Phys. Rev. B}\ }\textbf {\bibinfo {volume} {43}},\ \bibinfo
  {pages} {3706} (\bibinfo {year} {1991}{\natexlab{a}})}\BibitemShut {NoStop}%
\bibitem [{\citenamefont {Pan}\ and\ \citenamefont
  {Wang}(1991{\natexlab{b}})}]{pan1991linked}%
  \BibitemOpen
  \bibfield  {author} {\bibinfo {author} {\bibfnamefont {K.-K.}\ \bibnamefont
  {Pan}}\ and\ \bibinfo {author} {\bibfnamefont {Y.-L.}\ \bibnamefont {Wang}},\
  }\href@noop {} {\bibfield  {journal} {\bibinfo  {journal} {J. Appl. Phys.}\
  }\textbf {\bibinfo {volume} {69}},\ \bibinfo {pages} {4656} (\bibinfo {year}
  {1991}{\natexlab{b}})}\BibitemShut {NoStop}%
\bibitem [{\citenamefont {Thompson}\ \emph {et~al.}(1991)\citenamefont
  {Thompson}, \citenamefont {Yang}, \citenamefont {Guttman},\ and\
  \citenamefont {Sykes}}]{thompson1991high}%
  \BibitemOpen
  \bibfield  {author} {\bibinfo {author} {\bibfnamefont {C.}~\bibnamefont
  {Thompson}}, \bibinfo {author} {\bibfnamefont {Y.}~\bibnamefont {Yang}},
  \bibinfo {author} {\bibfnamefont {A.}~\bibnamefont {Guttman}}, \ and\
  \bibinfo {author} {\bibfnamefont {M.}~\bibnamefont {Sykes}},\ }\href@noop {}
  {\bibfield  {journal} {\bibinfo  {journal} {J. Phys. A}\ }\textbf {\bibinfo
  {volume} {24}},\ \bibinfo {pages} {1261} (\bibinfo {year}
  {1991})}\BibitemShut {NoStop}%
\end{thebibliography}
